\documentclass[a4paper,fleqn,usenatbib]{mnras}

\usepackage{newtxtext,newtxmath}

\usepackage[T1]{fontenc}
\usepackage{ae,aecompl}

\usepackage{graphicx}	
\usepackage{amsmath}	

\usepackage{adjustbox} 
\usepackage{lipsum}
\usepackage{caption} 
\usepackage[para,online,flushleft]{threeparttable} 

\usepackage{booktabs} 

\newcommand{\Me}{$M_\oplus$}

\newcommand{\rmin}{r_{\min}}
\newcommand{\rmax}{r_{\max}}

\newcommand{\amax}{a_{\max}}

\newcommand{\dgap}{\delta_\mathrm{g}}

\newcommand{\wgap}{w_\mathrm{g}}
\newcommand{\rgap}{r_\mathrm{g}}

\newcommand{\rmse}{e_\mathrm{rms}}
\newcommand{\rmsi}{i_\mathrm{rms}}

\newcommand{\es}{e_\mathrm{rms}^2}
\newcommand{\is}{i_\mathrm{rms}^2}

\title[On the sharpness of debris discs]{Constraining planetesimal
  stirring: how sharp are debris disc edges?}
\author[S. Marino]{ Sebastian Marino$^{1,2}$\thanks{E-mail:
    sebastian.marino.estay@gmail.com}
  \\
  $^{1}$Jesus College, University of Cambridge, Jesus Lane, Cambridge CB5 8BL, UK\\
  $^{2}$Institute of Astronomy, University of Cambridge, Madingley Road, Cambridge CB3 0HA, UK}
\date{Accepted XXX. Received YYY; in original form ZZZ}

\pubyear{2021}

\begin{document}
\label{firstpage}
\pagerange{\pageref{firstpage}--\pageref{lastpage}}
\maketitle

\begin{abstract}
  The dust production in debris discs by grinding collisions of
  planetesimals requires their orbits to be stirred. However, stirring
  levels remain largely unconstrained, and consequently the stirring
  mechanisms as well. This work shows how the sharpness of the outer
  edge of discs can be used to constrain the stirring levels. Namely,
  the sharper the edge is the lower the eccentricity dispersion must
  be. For a Rayleigh distribution of eccentricities ($e$), I find that
  the disc surface density near the outer edge can be parametrised as
  $\tanh[(r_{\max}-r)/l_{\rm out}]$, where $r_{\max}$ approximates the
  maximum semi-major axis and $l_{\rm out}$ defines the edge
  smoothness. If the semi-major axis distribution has sharp edges
  $\rmse$ is roughly $1.2 l_{\rm out}/r_{\max}$, or $\rmse=0.77 l_{\rm
    out}/r_{\max}$ if semi-major axes have diffused due to
  self-stirring. This model is fitted to ALMA data of five wide discs:
  HD~107146, HD~92945, HD~206893, AU~Mic and HR~8799. The results show
  that HD~107146, HD~92945 and AU~Mic have the sharpest outer edges,
  corresponding to $\rmse$ values of $0.121\pm0.05$,
  $0.15^{+0.07}_{-0.05}$ and $0.10\pm0.02$ if their discs are
  self-stirred, suggesting the presence of Pluto-sized objects
  embedded in the disc. Although these stirring values are larger than
  typically assumed, the radial stirring of HD~92945 is in good
  agreement with its vertical stirring constrained by the disc
  height. HD~206893 and HR~8799, on the other hand, have smooth outer
  edges that are indicative of scattered discs since both systems have
  massive inner companions.
\end{abstract}





\section{Introduction}
\label{sec:intro}

Planetesimal discs at tens of au analogous to the Kuiper belt are a
frequent component of planetary systems \citep{Su2006, Eiroa2013,
  Sibthorpe2018}. Mutual collisions within these belts grind down
solids producing high dust levels, which cause debris discs \citep[see
  reviews by][]{Wyatt2008, Krivov2010, Hughes2018}. This collisional
cascade requires planetesimal orbits to be dynamically excited or
\textit{stirred} by large planetesimals or planets, increasing their
eccentricities and inclinations (and thus their relative velocities)
enough such that mutual collisions lead to fragmentation. However,
even though the disc evolution through collisional processes has been
extensively studied analytically \citep[e.g.][]{Dominik2003,
  Wyatt2007hotdust, Lohne2008, Shannon2011, Pan2012, Geiler2017} and
via numerical simulations \citep[e.g.][]{Krivov2006, Thebault2007,
  Gaspar2012}, the stirring levels of debris discs remain largely
unknown.

Constraining the current eccentricity and inclination excitation
(i.e. stirring levels) of planetesimal discs is important since it
provides unique information on the dynamical history of planetary
systems. Proof of this is how the stirring or dynamical structure of
the Kuiper belt has been used to unveil the dynamical history of
Neptune \citep[see recent review by][]{Morbidelli2020}. For example,
the Kuiper belt's resonant population is direct evidence of Neptune's
radial migration in the past \citep{Malhotra1993, Malhotra1995}, and
its migration rate is well constrained by the inclination distribution
of the Kuiper belt's hot population \citep{Nesvorny2015}. Moreover,
the low inclinations of some detached and resonant Kuiper Belt objects
favours scenarios in which Neptune reached an eccentricity of
$\sim0.1$ during its migration \citep{Nesvorny2021}, possibly due to a
mild dynamical instability \citep[e.g. as in the Nice
  model,][]{Tsiganis2005}. Hence stirring levels can provide a very
complex picture of the dynamical history of planetary systems.

Determining the stirring level of planetesimal discs in exoplanetary
systems is, however, much more challenging because individual orbits
cannot be observationally determined. Instead, observations typically
constrain the projected surface brightness at multiple
wavelengths. Because the disc surface brightness depends on the dust
size distribution, orbital elements and dust temperatures, this is a
degenerate problem. Nevertheless, there are special cases in which
information on the orbital excitation of disc particles can be
directly obtained, especially at long wavelengths that are sensitive
to the emission of large grains whose dynamics are unaffected by
radiation pressure. Observations with the Atacama Large
Millimeter/submillimeter Array (ALMA) have been fundamental for this,
and below I summarise the special cases in which ALMA observations
have allowed constraining the excitation of eccentricities and/or
inclinations.

Recent ALMA observations of the edge-on disc around $\beta$~Pic
resolved its vertical extent \citep{Matra2019betapic}, and revealed a
vertical distribution that is inconsistent with a single Rayleigh
distribution of inclinations ($i$). Its vertical distribution was
better reproduced with two dynamical populations with typical
inclination dispersions ($\rmsi$) of $1\degr$ and $9\degr$. Such a
scenario is analogous to the classical Kuiper belt, which also has a
bimodal inclination distribution \citep[e.g.][]{Brown2001}. Similarly,
ALMA observations of AU~Mic marginally resolved its vertical extent
\citep{Daley2019} and constrained as well the distribution of
inclinations to $\rmsi\sim3\degr$.

For moderately inclined and narrow discs, the vertical extent can be
constrained by comparing the surface brightness as a function of
azimuth if the resolution is comparable or smaller than the disc width
\citep{Marino2016}. This is because the non-zero vertical thickness of
a disc widens the surface brightness distribution near the minor axis,
hence reducing slightly its surface brightness compared to a flat
disc. This principle has helped to marginally constrain the vertical
thickness of HR~4796's debris ring \citep{Kennedy2018}, finding
$\rmsi\sim3\degr$. Similarly, it is also possible to determine the
height of a wide disc if it contains narrow features such as a gap, as
done by \cite{Marino2019} for HD~92945's disc, finding
$\rmsi\sim4\degr$. Note that the latter case is very similar to
modelling done in protoplanetary discs with gaps to measure the degree
of dust settling \citep[e.g.][]{Pinte2016}.

The width of debris discs provides as well a direct upper limit on the
eccentricity levels. This is because for a uniform distribution of
longitudes of pericentres, the difference between pericentre and
apocentre distance must be equal or smaller than the disc width. This
is particularly interesting when applied to narrow debris discs
interacting with an eccentric planet since the final width of the disc
strongly depends on the forced eccentricity induced by the planet
\citep[e.g.][]{Pearce2014}. \cite{Kennedy2020} used these arguments to
show that the narrowness of the debris disc around Fomalhaut
\citep{MacGregor2017} and HD~202628 \citep{Faramaz2019} is
inconsistent with the standard secular evolution theory. The
dispersion of eccentricities in those discs, constrained by the disc
widths, is smaller than expected under the presence of an eccentric
planet. This inconsistency could be due to planetesimals being born on
eccentric and aligned orbits, or their relative velocities being
damped reducing the proper eccentricities \citep[e.g. through
  collisions,][]{Nesvold2013}.

The multiwavelength analysis of disc sizes can also constrain stirring
levels. By combining FIR and mm spatial information with collisional
evolution models, \cite{Geiler2019} found that the FIR emission of
HR8799's debris disc cannot be reproduced by a dynamically cold wide
belt that best matches the mm data. They concluded that to fit both
the large extention of the FIR emission and high mm flux, their model
required the presence of two populations: a dynamically cold and wide
disc and an excited disc with eccentricities of 0.3--0.5.

Similarly, mm observations that determine the disc emission spectral
index can constrain the grain size distribution
\citep[e.g.][]{Ricci2015, MacGregor2016, Lohne2020}. This is important
since the grain size distribution carries information about the
dynamical excitation of solids \citep[e.g.][]{Thebault2008, Pan2012},
and the strength of solids \citep{Obrien2003, Wyatt2011}. Therefore
constraining the mm spectral indices of discs could provide additional
constraints on the stirring levels. However, as pointed out by
\cite{Lohne2020}, the insufficient knowledge of the composition,
porosity and optical properties of solids in debris discs means that
estimates of the grain size distribution suffer from large systematic
uncertainties.

Finally, while observations at optical or NIR wavelengths can also
provide high fidelity characterisations of the disc surface brightness
in scattered light, using such observations to constrain the dynamical
excitation of the disc is even more challenging. This is because such
observations trace small grains at the bottom of the collisional
cascade, whose dynamics are strongly affected by radiation pressure
and stellar winds \citep[e.g.][]{Burns1979, Plavchan2005}. Such
processes increase the eccentricities of small grains once released
from larger bodies, and thus their dynamical stirring does not
directly trace the stirring of the larger bodies. Even their vertical
distribution is affected by radiation pressure and collisions
\citep{Thebault2009}, and thus observations that constrain the height
of small grains \citep[e.g.][]{Olofsson2020} do not necessarily trace
the vertical stirring of planetesimals. In principle, dynamical models
combined with radiative transfer simulations
\citep[e.g][]{Pawellek2019predictions, Pawellek201949ceti} could be
used to retrieve the stirring levels of the parent bodies. However, so
far this has proven very difficult due to degeneracies in model
parameters that set the radial distribution of small
grains. Furthermore, regardless of the planetesimal's stirring levels,
their bright halos that extend beyond the planetesimal discs are
expected to have surface brightness distributions following a
universal power law index of -3.5 \citep{Thebault2008}, erasing most
of the dynamical information of their parent bodies. Therefore,
observations at longer wavelengths that trace the distribution of
larger grains are better suited to constrain the stirring levels.

This paper focuses on how the eccentricity levels can be constrained
by studying how smooth or sharp the outer edge of a disc looks at long
wavelengths. While the outer edge smoothness depends both in the
semi-major axis distribution and level of eccentricities, the
smoothness places a direct upper limit on the eccentricity of
particles. In \S\ref{sec:model} I present a simple model of a stirred
disc that can be approximated by an analytic expression that
characterises the smoothness of the inner and outer edges due to
stirring. In \S\ref{sec:application} this model is used to fit ALMA
disc observations and constrain their eccentricity excitation. I
discuss the results in \S\ref{sec:dis} and I summarise the main
conclusions in \S\ref{sec:con}.












\section{A simple model of a stirred disc}
\label{sec:model}
This section aims to show how the surface density radial profile of a
disc changes depending on the level of orbital
excitation. Specifically, I will focus on the outer edge of a disc
which will become smoother for higher eccentricities and inclinations
of particles in a disc as they span a larger range of radii.

\begin{figure}
  \centering \includegraphics[trim=0.0cm 0.0cm 0.0cm 0.0cm,
    clip=true, width=1.0\columnwidth]{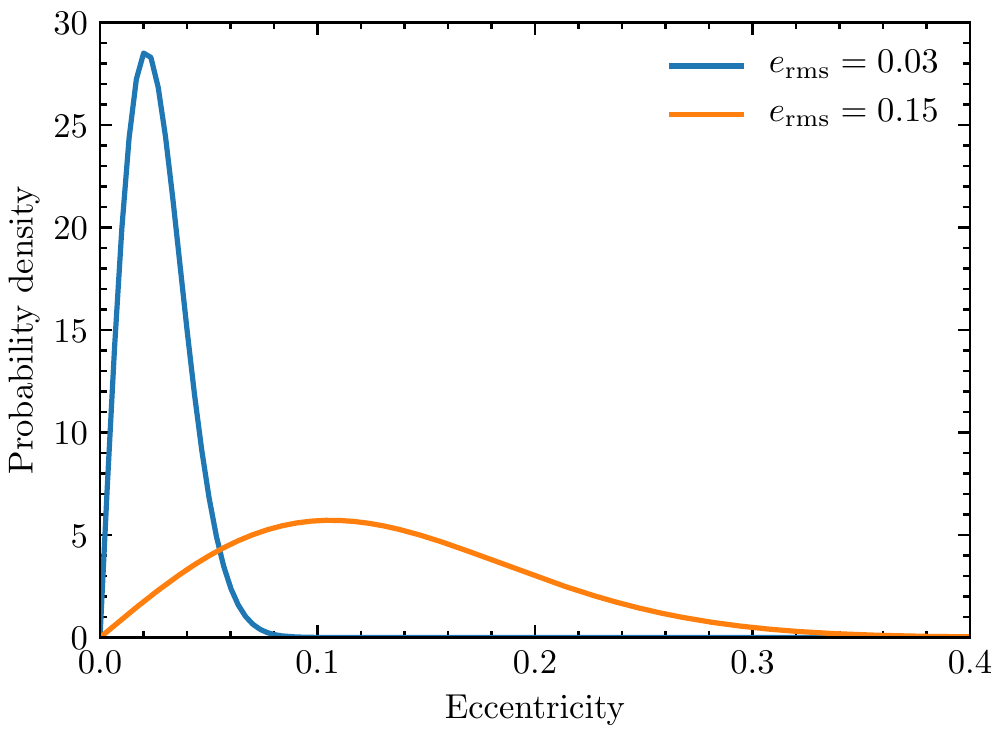}
  \caption{Rayleigh distributions of eccentricities with
    $\rmse=0.03$ in blue and 0.15 in orange.}
 \label{fig:dist_e}
\end{figure}

The disc is composed of an ensemble of particles in Keplerian orbits,
with randomised orbital parameters. Each particle has a different
semi-major axis ($a$) that is distributed as $N(a)da\sim
a^{\gamma+1}da$ between $a_{\min}$ and $a_{\max}$, with $\gamma$ being
the surface density \textit{slope} or power law index\footnote{The
surface density distribution is roughly proportional to
$N(a)/a$}. Eccentricities ($e$) and inclinations ($i$) follow a
Rayleigh distribution, i.e.
\begin{equation}
  f(e,i)=   \frac{4 ei}{\langle e^2\rangle \langle i^2\rangle} \exp\left[ - \frac{e^2}{\langle e^2 \rangle} - \frac{i^2}{\langle i^2\rangle}  \right],
\end{equation}
where $\langle e^2\rangle$ and $\langle i^2\rangle$ are the mean
square eccentricity ($\rmse^2$) and inclination ($\rmsi^2$). The choice of
a Rayleigh distribution is motivated by the expected orbital
distribution of ensembles of interacting planetesimals
\citep{Ida1992}. In addition, the dispersion of eccentricities and
inclinations is set to $\rmse=2\rmsi$, which is the expected
equilibrium when particles interact in a Keplerian potential
\citep{Ida1992, Ida1993ei}. The rest of the orbital parameters
(longitude of pericentre, longitude of ascending node and mean
anomaly), are considered to be uniformly distributed such that the
modelled discs are axisymmetric. Figure~\ref{fig:dist_e} shows an
example of a Rayleigh distribution for $e$ with $\rmse=0.03$ and
0.15. The mean is approximately $0.9\rmse$, the mode $0.7\rmse$ and
the standard deviation $0.5\rmse$.

\begin{figure}
  \centering \includegraphics[trim=0.0cm 0.0cm 0.0cm 0.0cm, clip=true,
    width=1.0\columnwidth]{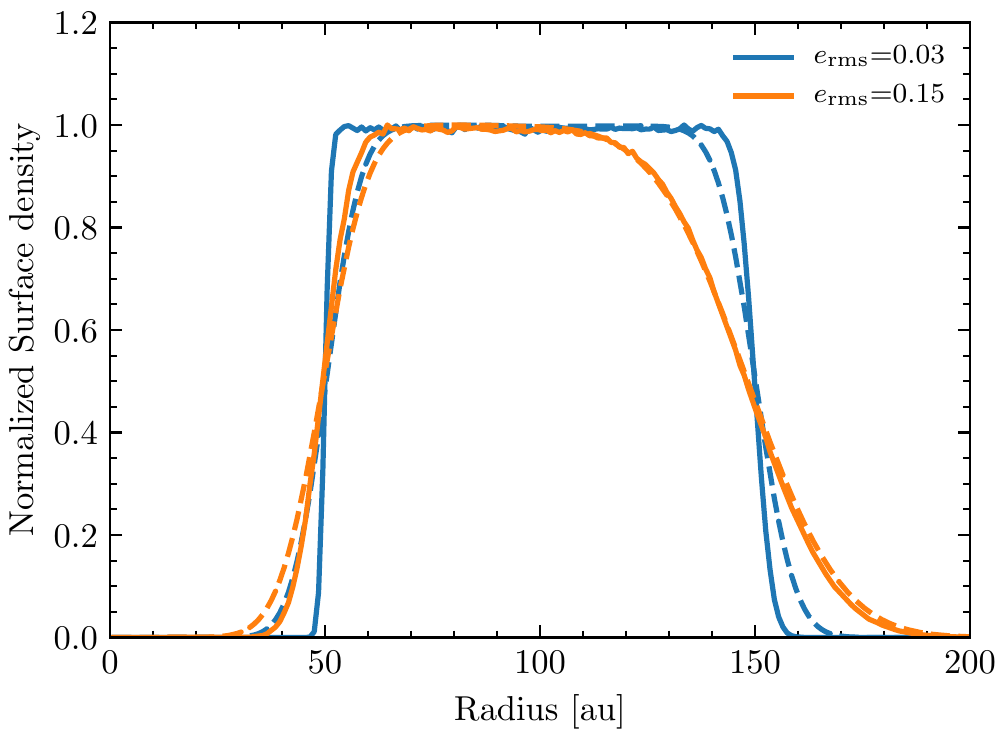}
  \caption{Normalised surface density of particles for eccentricity
    distributions with $\rmse=0.03$ in blue and 0.15 in orange. The
    solid lines represent the true surface densities, while the dashed
    lines are the result of convolving with a Gaussian kernel with a
    FWHM of 15~au.}
 \label{fig:Sr_conv}
\end{figure}

A total of $10^7$ orbits are drawn, from which the surface density is
computed as a function of the projected radius (hereafter referred to
simply as radius). Figure \ref{fig:Sr_conv} shows the surface density
for a disc with $\gamma=0$, $a_{\min}=50$, $a_{\max}=150$ and
$\rmse=0.03$ (blue) and 0.15 (orange). The solid lines correspond to
the surface density computed using 1~au wide radial bins. The dashed
lines correspond to the same profiles that are additionally convolved
with a Gaussian kernel/beam with a full width at half maximum (FWHM)
of 15~au or 10\% of $a_{\max}$. This is to mimic the response of an
interferometer such as ALMA. By comparing both solid and dashed lines,
some first conclusions can be drawn on the effect of particles
eccentricities. The higher $e$ is, the smoother the outer edge is as
particles spend time at a wider range of radii. Note that although
particles have a distribution of inclinations, the smoothing is
dominated by the eccentricities. This is because the radial smoothing
grows as $\mathcal{O}(e)$ and $\mathcal{O}(i^2)$, thus eccentricities
dominate. Another effect of higher $\rmse$ is that the radius at which
the density starts to drop is also shifted inwards. This is because
for large eccentricities the surface density at a given radius has
contributions from a larger range of semi-major axes, but for $r$
close to $a_{\max}$ this range is significantly smaller since no
particles have $a>a_{\max}$. Note that the inner edge also becomes
smoother with higher eccentricities, but this change is less
noticeable. This is because for a fixed eccentricity, the radial span
of an orbit is proportional to its semi-major axis, and thus at a
smaller radius the same fractional changes become less noticeable when
convolved with the same beam compared with the outer edge.


\begin{figure}
  \centering \includegraphics[trim=0.0cm 0.0cm 0.0cm 0.0cm, clip=true,
    width=1.0\columnwidth]{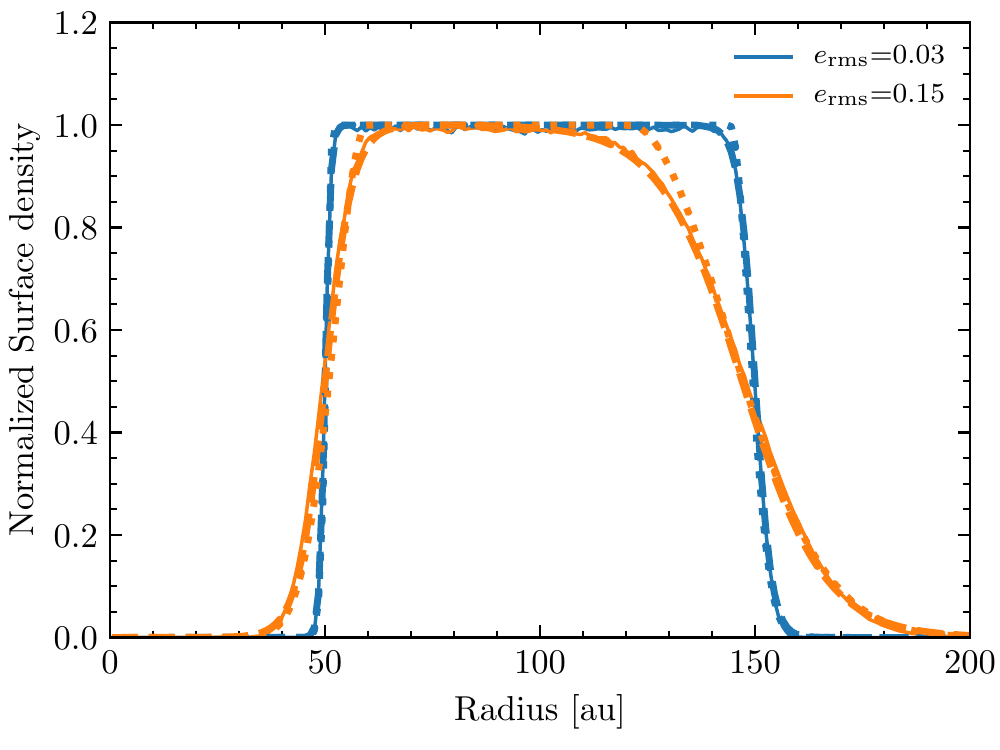}
  \caption{Normalised surface density of particles for eccentricity
    distributions with $\rmse=0.03$ in blue and
    0.15 in orange. The solid lines represent the true surface
    densities. The dashed and dotted lines represent the best fits
    when fitting the edges as Gaussian and hyperbolic tangent
    profiles, respectively. None of these models are convolved with a
    beam.}
 \label{fig:Sr_parmodels}
\end{figure}

I am interested in finding a parametric model that could reproduce the
smoothness of the inner and outer edges, and thus to constrain the
level of eccentricity by fitting such a model to disc observations. In
principle, it is also possible to fit distributions of orbital
elements \citep[e.g.][]{MacGregor2017, Kennedy2020}; however, this is
computationally expensive and thus much slower since typically
$\gtrsim10^6$ orbits must be drawn to obtain smooth models without
shot noise. I identify and test two potential simple models to
parametrise the edges: A model in which the edges have Gaussian
profiles and a model where the edges are approximated by hyperbolic
tangents. Figure \ref{fig:Sr_parmodels} compares the model surface
density obtained from randomised orbits and the parametric models that
best approximate to the surface density. The Gaussian edges (dotted
lines) produce smooth edges in the surface density profiles, but are
not smooth enough to reproduce the surface density near to where the
density starts to drop. On the other hand, the hyperbolic tangent
model reproduces better the profile across all radii for low and high
$\rmse$. Therefore, hereafter I adopt this parametrisation as default,
which is defined as
\begin{equation}
  \Sigma(r)=\frac{\Sigma_c}{4} \left(\frac{r}{r_{\min}}\right)^{\gamma} \left(1+\tanh\left[\frac{r-r_{\min}}{l_\mathrm{in}}\right]\right) \left(1+\tanh\left[\frac{r_{\max}-r}{l_\mathrm{out}}\right]\right), \label{eq:tanh}
\end{equation}
where $\gamma$ controls the surface density slope within the disc,
$r_{\min}$ and $r_{\max}$ approximate $a_{\min}$ and $a_{\max}$, and
$l_\mathrm{in}$ and $l_\mathrm{out}$ control the smoothness of the
inner and outer edges. Below in \S\ref{sec:retrieve} I show how the
ratio $l_\mathrm{out}/r_{\max}$ can be used to constrain $\rmse$.

\subsection{Retrieving $\rmse$}
\label{sec:retrieve}

\begin{figure}
   \centering \includegraphics[trim=0.0cm 0.0cm 0.0cm 0.0cm,
     clip=true, width=1.0\columnwidth]{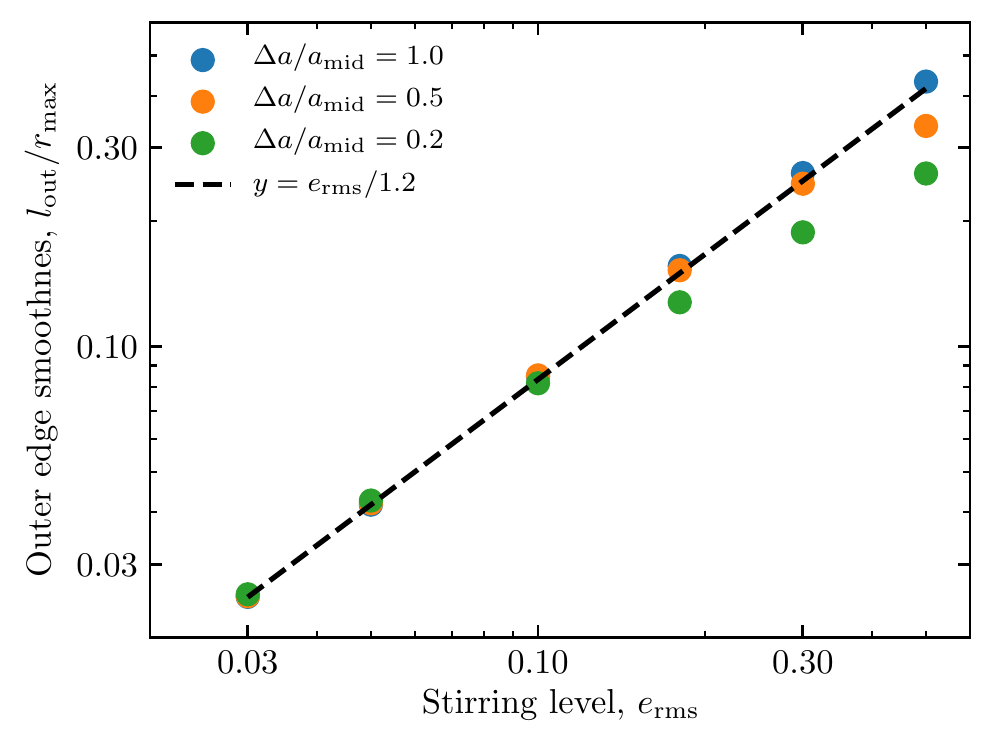}
  \caption{Best fit values for $l_\mathrm{out}/r_{\max}$ vs different
    input values of $\rmse$ when assuming a distribution of semi-major
    axes that has sharp edges. The colours represent different disc
    fractional widths in semi-major axis (i.e. width over the mid
    semi-major axis), with blue, orange and green representing
    fractional widths of 1, 0.5 and 0.2, respectively. The dashed line
    represents a linear relation between $l_\mathrm{out}/r_{\max}$ and
    $\rmse$ that approximates well the results, except for
    $\rmse>\Delta a/a_{\rm mid}$.}
 \label{fig:evsedr}
  \end{figure}

With a good analytical model in hand to reproduce the smoothness of
the disc edges, it is now important to test how well $\rmse$ can be
constrained by fitting such a model. I anticipate that the smoothness
of the disc outer edge here defined as $l_\mathrm{out}/r_{\max}$
should scale linearly with $\rmse$ for small values. In order to test
that I simulate the orbits of multiple discs with $\rmse=0.03, 0.05,
0.1, 0.18, 0.3, 0.5$, and fit them with the model defined by
Eq.~\ref{eq:tanh} leaving $\Sigma_c$, $r_{\min}$, $r_{\min}$,
$\gamma$, $l_{\rm in}$, and $l_{\rm out}$ as free
parameters. Figure~\ref{fig:evsedr} shows how the retrieved smoothness
defined by $l_\mathrm{out}/r_{\max}$ varies as a function of
$\rmse$. The different colours correspond to discs of different
fractional widths, here defined as the range of semi-major axes
divided by the mid semi-major axis. I find that the derived values for
$l_\mathrm{out}/r_{\max}$ correspond approximately to $\rmse/1.2$,
providing a simple relation to retrieve $\rmse$ when applying this
model to data.

Only when $\rmse$ becomes comparable to the fractional width, the
linear relation starts to break and $1.2 l_\mathrm{out}/r_{\max}$
underpredicts $\rmse$. In such cases the surface density is made of
particles from a wide range of radii, is strongly peaked, and it
resembles a Gaussian distribution. I find that for very narrow
semi-major axis distributions, the FWHM of the surface density
distribution is close to $1.7a_{\rm mid}\rmse$ (where $a_{\rm mid}$ is
the mid semi-major axis). This relation resembles what
\cite{Kennedy2020} found when studying eccentric and narrow debris
rings; namely, that the observed disc width is set by the orbital
excitation of particles in the disc.

\begin{figure}
   \centering
   \includegraphics[trim=0.0cm 0.0cm 0.0cm 0.0cm, clip=true,
    width=1.0\columnwidth]{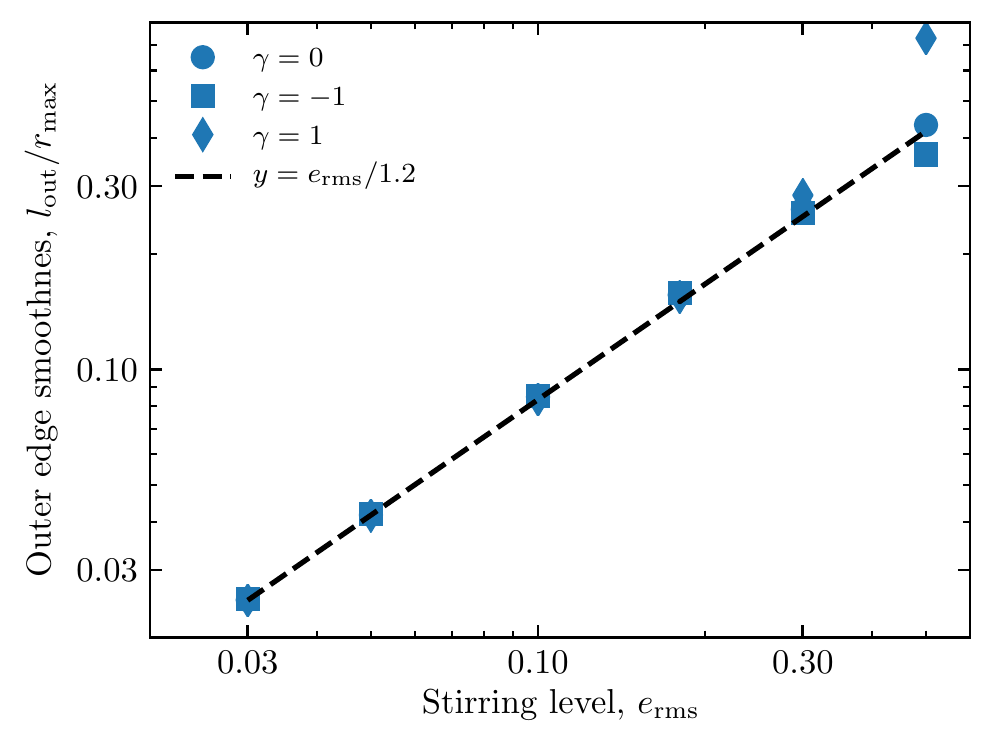}
  \caption{Best fit values for $l_\mathrm{out}/r_{\max}$ vs different
    input values of $\rmse$ when assuming a distribution of semi-major
    axes that has sharp edges and a fractional width of 1. The shape
    of the markers represent different slopes of the surface density
    distribution, with circles, squares and diamonds representing
    power law indices of 0, -1 and 1, respectively. The dashed line
    represents a linear relation between $l_\mathrm{out}/r_{\max}$ and
    $\rmse$ that approximates well the results.}
 \label{fig:evsegamma}
\end{figure}

I also test how well $\rmse$ can be retrieved when the surface density
slope is different than zero and $\gamma$ is still left as a free
parameter. Figure~\ref{fig:evsegamma} shows $l_\mathrm{out}/r_{\max}$
as a function of $\rmse$ for $\gamma=-1,0,1$. I find that $1.2
l_\mathrm{out}/r_{\max}$ still approximates $\rmse$ well up to values
of $\sim0.5$, at which point $\rmse$ would be overestimated for
$\gamma=-1$ and slightly underestimated for $\gamma=1$. Therefore I
conclude that the chosen parametric model is robust in retrieving the
value of $\rmse$, except when the fractional width of semi-major axes
becomes comparable to $\rmse$.


A caveat in the analysis presented above is that it assumes that the
distribution of semi-major axes has a sharp cut, and thus any
smoothness in the surface density profile is attributed primarily to
the eccentricity of particles. While this might not be the case, the
scenario assumed here can be used to set an upper limit in the
eccentricity of particles in a disc. This is because higher
eccentricities would make the outer edge appear even smoother for any
distribution of semi-major axes. Therefore, the sharp cut in
semi-major axes beyond $a_{\max}$ is the best case scenario to produce
a sharp edge. Any other continuous distribution of $a$ would make the
outer edge even smoother. Therefore this model offers a useful tool to
at least set an upper limit on $\rmse$.

\subsection{Smooth edges in the semi-major axis distribution}
\label{sec:smootha}
As stated above, so far I have assumed that the semi-major axis
distribution has sharp edges; however, there are scenarios in which
the edges in the semi-major axis distribution of dust or planetesimals
could become smoother with time. This could be expected in a
self-stirring scenario, in which the increase in eccentricities due to
close encounters leads to a diffusion in semi-major axis as well
\citep[see e.g. Fig 3 in][]{Ida1993}. Assuming the pericentre distance
between encounters is kept constant, particles stirred to
eccentricities $e$ diffuse in semi-major axis by a length $\sim
ae/(1-e)$ \citep[e.g.][]{Duncan1987}. If this is the case, the
observed smoothness in the surface density would be a combination of
both eccentricities and the smooth semi-major axis distribution.

\begin{figure}
   \centering \includegraphics[trim=0.0cm 0.0cm 0.0cm 0.0cm,
     clip=true, width=1.0\columnwidth]{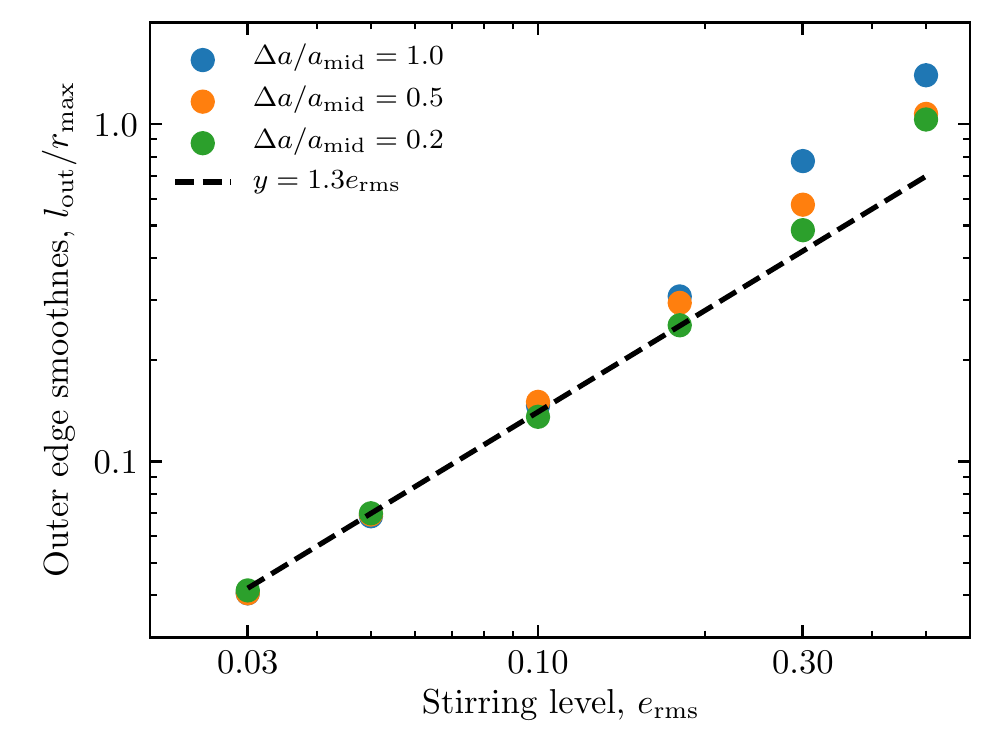}
  \caption{Best fit values for $l_\mathrm{out}/r_{\max}$ vs different
    input values of $\rmse$ when assuming a distribution of semi-major
    axes that have smooth edges. The colours represent different disc
    fractional widths in semi-major axis (i.e. width over the mid
    semi-major axis), with blue, orange and green representing
    fractional widths of 1, 0.5 and 0.2. The dashed line represents a
    linear relation between $l_\mathrm{out}/r_{\max}$ and $\rmse$ that
    approximates well the results, except for broad discs with high
    $\rmse$ values.}
 \label{fig:evsedr_da}
  \end{figure}

In order to test the effect of both effects combined, new calculations
similar to those shown above in \S\ref{sec:model} are performed. Now
instead of having a semi-major axis distribution $N(a)$ following a
simple power law distribution, now $N(a)$ is defined as a power law
with smooth edges (assuming the same functional form as Equation
\ref{eq:tanh}). Namely
\begin{equation}
  N(a)\propto \left(\frac{a}{a_{\min}}\right)^{\gamma+1} \left(1+\tanh\left[\frac{a-a_{\min}}{s_\mathrm{in}}\right]\right) \left(1+\tanh\left[\frac{a_{\max}-a}{s_\mathrm{out}}\right]\right), \label{eq:Na_smooth}
\end{equation}
where the smoothness parameters $s_{\rm in}$ and $s_{\rm out}$ are set
to $a_{\min,\max}\rmse/(1-\rmse)$ at the inner and outer edge,
respectively. This additional smoothing means that for the same
stirring level $\rmse$, the edges of a disc are smoother than in the
case considered in \S\ref{sec:model}. With this additional semi-major
axis smoothing I fit the same parametric model defined by Equation
\ref{eq:tanh}, and find that the best fit value of
$l_\mathrm{out}/r_{\max}$ now corresponds to approximately $1.3\rmse$
for $\rmse\lesssim0.2$ as shown in Figure \ref{fig:evsedr_da}. Hence
the same observed level of smoothness corresponds to a lower degree of
orbital excitation in this scenario. When discussing the observed
smoothness levels in discs and its connection to self-stirring in
\S\ref{sec:selfstirring}, I will consider this scenario to interpret
the results.

Finally, I also find that the derived values of $r_{\min}$ and
$r_{\max}$ approximate well $a_{\min}$ and $a_{\max}$ ($<10\%$ error)
for $\rmse\lesssim0.2$ when assuming smooth edges in the semi-major
axis distribution, and for $\rmse\lesssim0.5$ when assuming sharp
edges in the semi-major axis distribution. Therefore, $r_{\min}$ and
$r_{\max}$ are also good proxies for determining the semi-major axis
distribution, and $2(r_{\max}-r_{\min})/(r_{\max}+r_{\min})$
approximates the fractional width of semi-major axes.


\subsection{Required resolution and S/N}


In order to measure how well $\rmse$ can be retrieved from a noisy
radial profile, I produce multiple models with different resolutions
(20\%, 10\% and 5\% of $\amax$) by convolving with a Gaussian beam,
different signal-to-noise ratios (SNR) per beam\footnote{Note that
this is the SNR of the radial profile. This is always higher than the
SNR in reconstructed images since it is obtained by averaging the
surface brightness over azimuthal angle or by fitting an axisymmetric
model to the data.}  (15, 30 and 60), and different $\rmse$ values
(0.025, 0.05, 0.1, 0.2, 0.4). I fit the parametric model convolved
with the same Gaussian beam to these mock profiles using an MCMC
procedure \citep{emcee} and estimate $\rmse$ as $1.2
l_\mathrm{out}/r_{\max}$. I find that the fractional error depends on
the beam size and SNR as follows
\begin{equation}
\frac{\delta \rmse}{\rmse} \approx 0.1 \left(\frac{60}{\rm
  SNR}\right)\left(\frac{\mathrm{beam}\ d/a_{\max}}{0.1}\right),
\end{equation}
where $d$ is the distance in pc, beam is the beam size in arcseconds
and $a_{\max}$ is the maximum semi-major axis. If $\rmse\ll
\left(\frac{\mathrm{beam}\ d/a_{\max}}{0.1}\right)$, then $\rmse$ is
not well constrained but a 99.7\% confidence upper limit can be set
which approximates to
\begin{equation}
 \max(\rmse)\approx 0.03 \left(\frac{60}{\rm
  SNR}\right)\left(\frac{\mathrm{beam}\ d/a_{\max}}{0.1}\right).
\end{equation}

Therefore, observations that have resolved well the disc emission
(with a few beams across) and with SNR$\gtrsim20$, can be used to set
meaningful constraints on $\rmse$. In \S\ref{sec:application} I fit
the parametric model described by Equation \ref{eq:tanh} to ALMA
observations of 5 discs, and based on the derived smoothness of their
outer edges I constrain their eccentricity levels considering the two
models presented above, i.e. sharp and smooth semi-major axes
distributions.

\section{Application to ALMA data}
\label{sec:application}

\begin{figure*}
  \centering \includegraphics[trim=0.0cm 0.0cm 0.0cm 0.0cm, clip=true,
    width=0.9\textwidth]{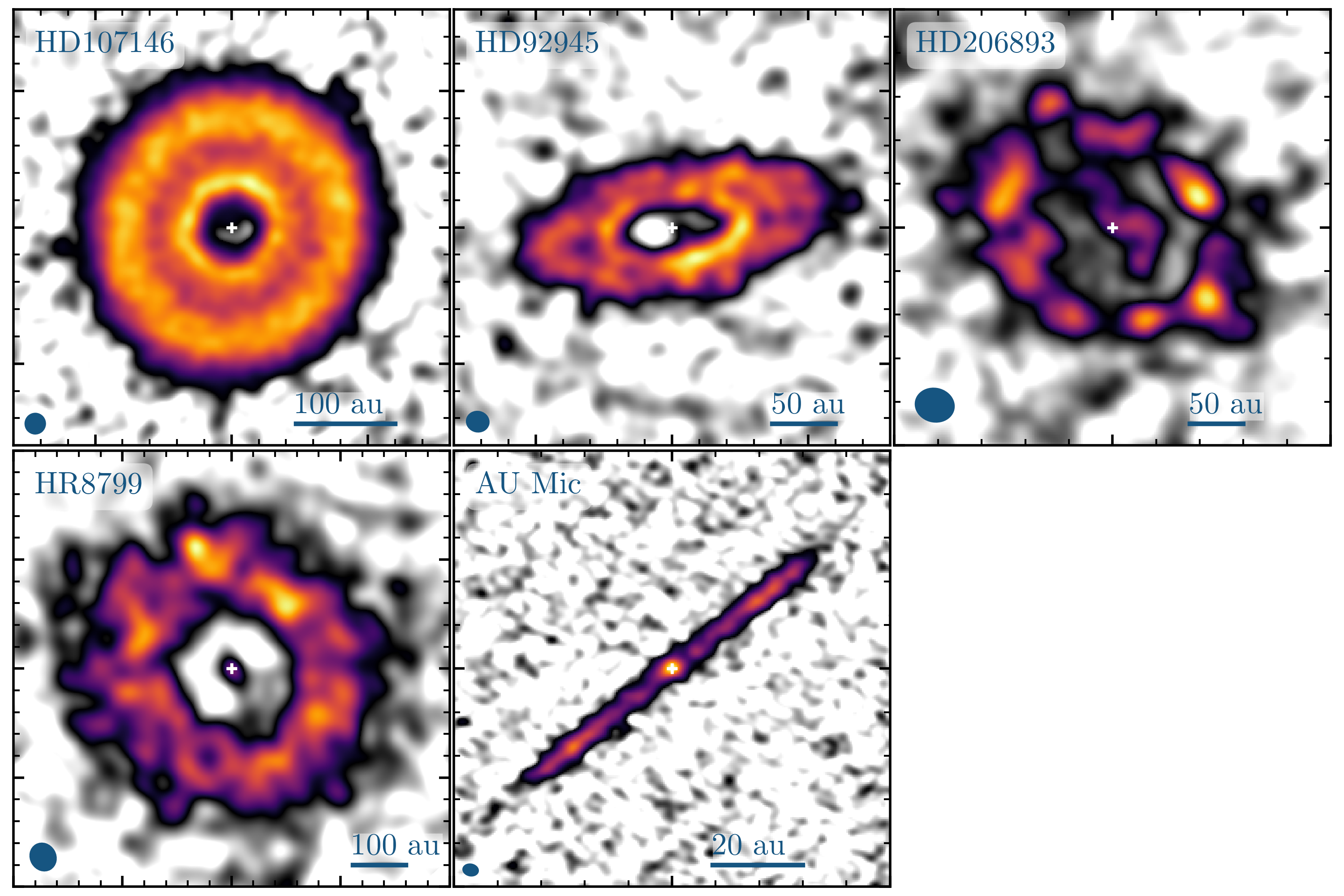}
  \caption{Dust continuum ALMA images of the five debris discs studied
    here. The images were obtained using the \textit{clean} algorithm
    implemented in the \textsc{tclean} task in \textsc{CASA}. Only
    values above a significance level above $3\sigma$ are shown in
    colour. The beams are represented as blue ellipses at the bottom
    left corner in each panel, with sizes of (from left to right)
    $0\farcs80\times0\farcs79$, $0\farcs87\times0\farcs79$ and
    $0\farcs92\times0\farcs80$ in the top row, and
    $1\farcs33\times1\farcs27$ $0\farcs42\times0\farcs33$ in the
    bottom row. These images are not primary beam corrected, and thus
    the noise levels are constant across the images. HD~107146's image
    (top left) is obtained from 1.14~mm data after subtracting the
    best fit of a background source near the disc inner edge, and
    using Briggs weighting (robust=0.5). HD~92945's image (top middle)
    is obtained from 0.86~mm data using natural weights and a uvtaper
    of $0\farcs7$. HD~206893's image (top right) is obtained from
    1.35~mm data using Briggs weighting (robust=2.0) and a uvtaper of
    $0\farcs4$. HR~8799's image (bottom left) is obtained from 0.88~mm
    data after subtracting the best fit of a background source near
    the disc inner edge, and using natural weights and a uvtaper of
    $0\farcs8$. AU~Mic's image (bottom middle) is obtained from
    1.35~mm data using Briggs weighting (robust=0.5). The white cross
    in the middle of the images represents the stellar position. The
    ticks in all panels are spaced by $1\arcsec$. A scale bar is shown
    at the bottom right of each panel.}
 \label{fig:images}
\end{figure*}

I wish to apply the model presented above to real data to constrain
the level of stirring of some planetesimal discs. The best data to do
this are ALMA observations of discs that have resolved their widths
with multiple resolution elements (or beams) with SNR~$>10$. I
identify 5 disc observations as the most promising for this exercise:
HD~107146 \citep{Marino2018hd107}, HD~92945 \citep{Marino2019},
HD~206893 \citep[][]{Marino2020hd206, Nederlander2021}, AU~Mic
\citep{Daley2019} and HR~8799 (Faramaz et al. submitted). $\beta$~Pic
would also be a good candidate since it is very well resolved with
ALMA. However, it is known to be asymmetric, warped, and possibly has
multiple dynamical components \citep[][and references
  therein]{Matra2019betapic}. Thus it is excluded from this
sample. Figure \ref{fig:images} shows ALMA continuum images of the
five discs analysed here.

\subsection{Fitting procedure}

The observed visibilities of each disc are fitted by first creating
synthetic images using \textsc{RADMC3D} \citep{radmc3d} of the dust
continuum emission. For this, a dust surface density defined as
Equation~\ref{eq:tanh} is input into \textsc{RADMC3D}, together with a
stellar model spectrum consistent with previous studies of these
systems. Because HD~107146, HD~92945, HD~206893 and AU~Mic are known
to have a gap in between a broad disc \citep[AU~Mic's disc was
  described instead as double belts by][]{Daley2019}, their models
include a Gaussian gap to reproduce the surface density as best as
possible. In cases where multi band data are available (HD~107146,
HD~206893 and HR~8799), only one image is produced with
\textsc{RADMC3D} and the rest obtained by scaling the disc and central
emission. The disc emission is scaled using a disc spectral index,
$\alpha_\mathrm{mm}$, which I leave as a free parameter. The stellar
emission in the central pixel is also scaled assuming a spectral index
of 2.

I further assume a vertical mass distribution that is Gaussian
\citep[consistent with a Rayleigh distribution of
  inclinations,][]{Matra2019betapic}, with a standard deviation that
scales linearly with radius as $hr$. Given the low inclination of
HD~107146, HD~206893 and HR~8799, $h$ is hardly constrained and thus I
simply fix it to 0.05. For HD~92945 and AU~Mic, $h$ is left as a free
parameter since these two discs are seen at a higher inclination and
thus $h$ can be constrained \citep{Marino2019, Daley2019}.

Each synthetic image is then multiplied by the corresponding antenna
primary beam, which is different for different wavelengths and antenna
arrays (either 12m or ACA). Model visibilities are obtained based on
those images using a Fast Fourier Transform algorithm, and a bilinear
interpolation to extract visibilities at the same $u-v$ points as the
observations \citep[as in][]{Marino2018hd107}. Additionally, I
consider phase center offsets for each data set, which are applied by
multiplying each model visibility by $\exp[-2\upi i
  (u\Delta\mathrm{RA}+v\Delta_\mathrm{DEC})]$, where
$\Delta\mathrm{RA}$ and $\Delta\mathrm{DEC}$ are the sky offsets in
right ascension and declination directions, respectively. Finally the
$\chi^2$ is computed for each set of parameters, and this is used to
feed an Affine Invariant MCMC Ensemble sampler using \textsc{emcee}
\citep{GoodmanWeare2010, emcee} to estimate the posterior distribution
of each parameter.

It is well known that the estimated uncertainties or weights of the
observed visibilities delivered by the ALMA calibration pipeline are
often off by a small factor, even after using tools such as
\textsc{statwt} that determine the visibility dispersion with CASA
\citep{casa}. To correct for this I multiply the weights of each data
set by a fixed value to force the reduced $\chi^2$ ($\chi^2_{\nu}$) to
be approximately 1 as in \cite{Marino2018hd107}. Because the SNR of
each visibility is $\ll1$, these fixed values can be found with a null
model. A subtle but important detail here is that $\chi^2_{\nu}$ is
defined as $\chi^2$ divided by the number of data
points\footnote{Formally $\chi^2_{\nu}$ is defined as $\chi^2$ divided
by the degrees of freedom. In this case this is almost equal to the
number of data points which are typically $\sim10^6$ for ALMA 1h
integrations and moderate time and spectral averaging.}, which in this
case is twice the number of observed visibilities since each
visibility has a real and imaginary component that are
independent. This was not realised in \cite{Marino2018hd107,
  Marino2019} and thus the reported uncertainties were overestimated
by a factor $\sqrt{2}$.

To conclude, each model is fitted with at least 10 parameters that set
the system orientation, phase center offsets and surface
density. Additional parameters are used to fit the disc height,
spectral indices, sub-millimeter galaxies (SMGs), gaps in the discs
and central unresolved flux. In appendix \ref{appendix:fit} I describe
a few particularities in the fitting procedure of each disc. The
results of the fits are presented below.



\subsection{Results}


\begin{table*}
  \centering
  \caption{Retrieved fractional width of semi-major axes,
    $\rmse$ and $\rmsi$. The fractional width is defined as
    $2(\rmax-\rmin)/(\rmax+\rmin)$ (2$^{\rm nd}$ column). The values
    for $\rmse$ are estimated as $1.2 l_\mathrm{out}/r_{\max}$ when
    assuming a semi-major axis distribution with sharp edges (3$^{\rm
      rd}$, 4$^{\rm th}$ and 5$^{\rm th}$ column), or as $0.77
    l_\mathrm{out}/r_{\max}$ when assuming a semi-major axis
    distribution with smooth edges (6$^{\rm rd}$, 7$^{\rm th}$ and
    8$^{\rm th}$ column). The values for $\rmsi$ are estimated as
    $\sqrt{2}h$ (9th column). The best fit values correspond to the
    median, with uncertainties based on the 16th and 84th percentiles
    of the marginalised distributions. The lower and upper limits are
    based on the 5\% and 95\% percentiles}
  \label{tab:results}
  \begin{adjustbox}{max width=1.0\textwidth}
    \begin{tabular}{lcccc c c c c } 
  \hline
  \hline
  System & $\Delta r/r$ & \multicolumn{3}{c}{$\rmse$ ($N(a)$ with sharp edges)}  & \multicolumn{3}{c}{$\rmse$ ($N(a)$ with smooth edges)}  & $\rmsi$ \\
  (1) & (2) & (3) & (4) & (5) & (6) & (7) & (8) & (9)  \\
  \cmidrule(lr){1-1}\cmidrule(lr){2-2}\cmidrule(lr){3-5}\cmidrule(lr){6-8} \cmidrule(lr){9-9} 
  HD~107146 & $1.13\pm0.01$ & $0.18\pm0.01$ & >0.16  & <0.20  & $0.121\pm0.005$ & >0.11  & <0.13 &  \\
  HD~92945 & $0.87^{+0.05}_{-0.09}$ & $0.23_{-0.08}^{+0.11}$ & >0.11 & <0.43 & $0.15_{-0.05}^{+0.07}$ & >0.07  & <0.28 & $0.061\pm$0.020 \\
  HD~206893 & $1.20^{+0.42}_{-0.26}$ & $0.49_{-0.12}^{+0.11}$ & >0.30  & <0.67  & $0.31_{-0.08}^{+0.07}$ & >0.19  & <0.43 &   \\
  HR~8799  &  $0.73_{-0.22}^{+0.11}$ & $0.55_{-0.19}^{+0.24}$ & >0.25 & <0.99  & $0.35_{-0.12}^{+0.16}$ & >0.16 & <0.63 & \\
  AU~Mic  &  $0.97_{-0.22}^{+0.41}$ & $0.16_{-0.04}^{+0.03}$ & >0.10  & <0.22 & $0.10\pm0.02$ & >0.06  & <0.14 & $0.021\pm$0.004 \\
  \hline
  \end{tabular}
  
  \end{adjustbox}
\end{table*}

\begin{figure}
  \centering \includegraphics[trim=0.0cm 0.0cm 0.0cm 0.0cm, clip=true,
    width=1.0\columnwidth]{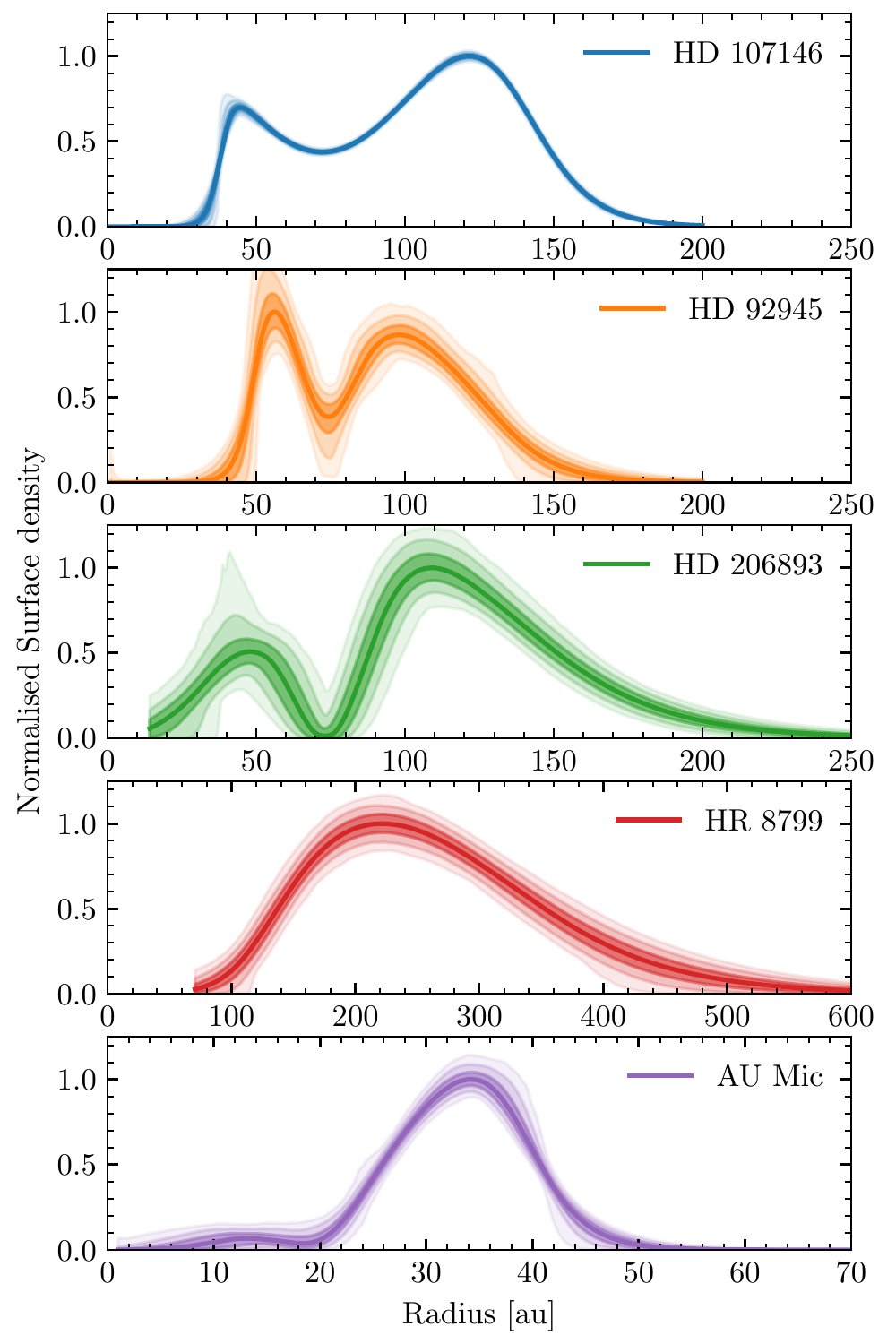}
  \caption{Posterior distribution of the surface density distribution
    for the five discs analysed in this work. The solid line
    represents the median, while the shaded areas represent the 68, 95
    and 99.7\% confidence regions. Note that these are parametric
    model profiles that that best adjust to the visibilities, and thus
    are not obtained from any imaging procedure and are not convolved
    with any beam. Similarly, these profiles do not depend on the
    assumption of sharp or smooth edges in the semi-major axis
    distribution.}
 \label{fig:rad}
\end{figure}

Table \ref{tab:results} presents the retrieved fractional widths and
$\rmse$, and Figure \ref{fig:rad} shows the posterior distribution of
the model surface density profiles. The best fit values for the disc
parameters are presented in Table~\ref{tab:mcmc}. The first immediate
result is that the five discs are very broad with high fractional
widths. This is not surprising since the sample only includes discs
that are resolved with multiple beams across their width, naturally
biasing this sample. However, HD~206893, HR~8799 and AU~Mic have high
uncertainties in their fractional widths. These large uncertainties
are due to the degeneracies between $\gamma$ and $r_{\min}$ and
$r_{\max}$. Nevertheless the true radial extent of the discs is better
constrained by the model radial profiles shown in Figure
\ref{fig:rad}. These clearly show the known gaps in HD~107146,
HD~92945 and HD~206893, a broad and smooth disc around HR~8799, and
low level emission extending down to $\sim10$~au for AU~Mic supporting
the conclusions of \cite{Daley2019}. Whether this inner emission is
separated from the emission further out by a gap is less clear in this
system.


The second immediate result is that the five discs have outer edges
that are best fit by a smooth transition, which could be indicative of
a significant level of stirring as shown in \S\ref{sec:model}. Table
\ref{tab:results} presents the inferred $\rmse$ (median values and
lower and upper limits) when assuming a semi-major axis distribution
($N(a)$) with sharp edges (3rd, 4th and 5th columns) or smooth edges
(6th, 7th and 8th column). The sharpest surface density outer edges
are found around HD~107146, HD~92945 and AU~Mic, with estimated
$\rmse$ values of $0.18\pm0.01$, $0.23^{+0.11}_{-0.08}$ and
$0.16^{+0.03}_{-0.04}$, respectively, assuming sharp edges for
$N(a)$. When considering a self-stirred case, i.e.  $N(a)$ having
smooth edges, the outer edges of these three discs suggest $\rmse$
values of $0.121\pm0.005$, $0.15^{+0.07}_{-0.05}$ and
$0.10\pm0.02$. Whether these values truly correspond to the level of
stirring in these three discs is uncertain, but they offer at least an
upper limit to the stirring levels in these discs.

A particular test can be done by comparing the derived values of
$\rmse$ and $h$ since it is expected that $\rmse\sim2\rmsi$ if discs
are in equilibrium, and $h$ is directly related to $\rmsi$. As shown
by \cite{Matra2019betapic} for a Rayleigh distribution of
eccentricities $\rmsi=\sqrt{2}h$, and thus $\rmse=\sqrt{8}h$ if these
discs are in equilibrium. For HD~92945 and AU~Mic, $h$ was left as a
free parameter since it can be constrained given the high inclination
of these discs from face on orientation. The results show that the
vertical aspect ratios are constrained to values of $0.043\pm0.014$
and $0.015\pm0.003$ for these discs, respectively.  Thus, given the
vertical extent of HD~92945 and AU~Mic, the expected values for
$\rmse$ are $0.12\pm0.04$ and $0.04\pm0.008$, respectively. Thus, the
condition $\rmse\sim2\rmsi$ is met for HD~92945, with inferred levels
of vertical and radial stirring that are consistent with being in
equilibrium. This is not the case for AU~Mic for which a significantly
larger value of radial stirring is found, which is in a $3\sigma$
tension. Three reasons are identified for why this could be the
case. First, in a self-stirred disc the eccentricities grow much
faster than the inclinations \citep{Ida1993, Krivov2018stirring} which
could explain the low $h$ and high $\rmse$ in AU~Mic since this system
is very young \citep[$\sim18.5^{+2.0}_{-2.4}$~Myr,][]{Miret-Roig2020}
compared with HD~92945 \citep[120--250~Myr,][]{Mesa2021}. Second, $h$
might be underestimated as \cite{Daley2019} found a value for $h$
($0.027^{+0.004}_{-0.005}$) that is 30\% larger than the best fit
presented here, which would solve the tension. This difference on the
estimated $h$ could be due to differences in the chosen parametric
model. Third, it could be that its semi-major axis distribution has an
outer edges that is smoother than assumed here, and thus the true
value of $\rmse$ is lower and closer to 0.04.


The eccentricity values found for HD~206893 and HR~8799 are much
larger and this can be seen directly from their smoother outer edges
in the best fit model radial profiles in Figure~\ref{fig:rad}. These
two systems host massive companions that have been directly imaged,
one being likely a brown dwarf at 11~au around HD~206893
\citep{Milli2017, Delorme2017, Grandjean2019, Ward-Duong2021} and four
giant planets around HR~8799 \citep{Marois2008, Marois2010}. The
formation, evolution and interaction of these companions with the disc
could explain the particularly high eccentricities or extended outer
edges. In fact, as proposed by \cite{Geiler2019}, HR~8799's
multiwavelength disc emission is best fit as the combination of a
dynamically cold and hot populations. The latter could be a scattered
disc of particles having close encounters with HR~8799~b that has a
semi-major axis of $\sim70$~au \citep{Wang2018}, or a resonant
population trapped by HR~8799~b as it migrated in the past. In these
two scenarios, the assumption about a single Rayleigh distribution of
eccentricities and inclination breaks down and the derived values are
not longer representative of the true eccentricities. Scenarios of
planet disc interactions are further discussed in
\S\ref{sec:planetstirring}.

\begin{figure}
  \centering \includegraphics[trim=0.0cm 0.0cm 0.0cm 0.0cm, clip=true,
    width=1.0\columnwidth]{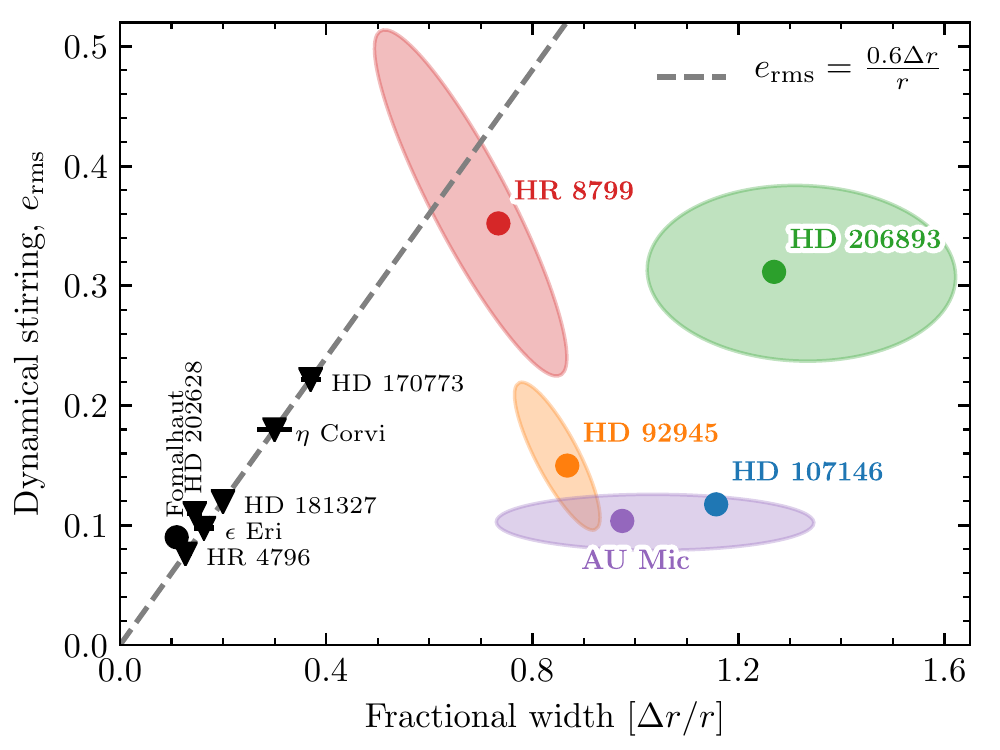}
  \caption{Inferred dynamical stirring $\rmse$ vs disc fractional
    width of the 5 broad discs analysed in this work (coloured
    symbols) and a sample of 6 narrow discs \citep[black symbols,
      $\epsilon$~Eri, $\eta$~Corvi, HR~4796, HD~202628, HD~170773,
      Fomalhaut, and HD~181327,][]{Booth2017epseri,
      Marino2017etacorvi, Kennedy2018, Faramaz2019, Sepulveda2019,
      Kennedy2020, Pawellek2021}. The $\rmse$ values for HD~107146,
    HD~92945, HD~206893, HR~8799 and AU~Mic are based on the their
    outer edges and assume a smooth edge in their semi-major axis
    distribution (i.e. self-stirred scenario). The $\rmse$ values for
    Fomalhaut and HD~202628 correspond to the dispersion of proper
    eccentricities based on fitting orbital distributions fits to the
    observed surface brightness. The $\rmse$ upper limits for HR~4796,
    $\epsilon$~Eri, HD~170773, HD~181327 and $\eta$~Corvi are set
    equal to half their fractional widths, i.e. $\rmse\leq0.6\Delta
    r/r$ (grey dashed line). This latter limit is only strictly valid
    for a circular discs. The coloured dots and ellipses represent the
    medians and $1 \sigma$ confidence intervals based on the posterior
    distributions. Note that the ellipse of HD~107146 is smaller than
    the blue dot and this is not visible. }
 \label{fig:summary}
\end{figure}

Figure~\ref{fig:summary} summarises the derived $\rmse$ of the five
discs analysed here as a function of their fractional
width\footnote{The derived fractional width of these five discs does
not necessarily correspond to the fractional width of the surface
density distribution defined by its FWHM, especially for cases with
$\gamma$ different from zero and high eccentricities. The fractional
width defined by $2(\rmax-\rmin)/(\rmax+\rmin)$ better represents the
fractional width of semi-major axes.}, and under the assumption of a
semi-major axis distribution with smooth edges. The grey dashed line
shows the maximum $\rmse$ for a circular belt, which corresponds to a
case of a narrow semi-major axis distribution where the the belt FWHM
is set entirely by planetesimal eccentricities ($\Delta r=1.7r\rmse$,
\S\ref{sec:retrieve}). The same figure shows the derived proper
eccentricity dispersion of the narrow and eccentric belts around
Fomalhaut and HD~202628 \citep{Kennedy2020}\footnote{In a disc that is
eccentric due to a forced eccentricity, the dispersion of
eccentricities that set the relative velocities depends both on the
mean and standard deviation of proper eccentricities. Since in
\cite{Kennedy2020} the derived values or upper limits of the standard
deviation are larger than for the mean, I consider the standard
deviation of proper eccentricities as a better representation of
$\rmse$.}, and upper limits of $\rmse$ for the narrow belts around
HR~4796, $\epsilon$~Eri, HD~170773, HD~181327 and $\eta$~Corvi based
on their fractional widths ($\rmse\leq0.6\Delta r/r$). This comparison
illustrates how the eccentricity values of wide discs derived here are
lower than the most basic upper limit based on their widths, and
comparable with the upper limits of some of the narrowest
discs. Future ALMA observations that resolve the width of both narrow
and wide discs with multiple beams could constrain the outer edge
smoothness of an even larger sample.

Note that the very smooth outer edge of HR~8799's appears to be
inconsistent with the maximum $\rmse$ represented by the dashed
line. This apparent inconsistency is mainly due to two effects. First,
the eccentricity values for HR~8799 and HD~206893 are likely
overestimated. Figure~\ref{fig:evsedr_da} shows that for $\rmse>0.2$,
the smoothness is greater than the linear regression. This means that
the derived eccentricity for a very smooth edge will be overestimated,
and thus the values derived for HR~8799 and HD~206893 must be taken
with caution. Second, the width of HR~8799's disc
($r_{\max}-r_{\min}$) does not correspond necessarily to its
FWHM. This is especially true for $\gamma$ values different from zero,
in which case $r_{\max}-r_{\min}$ approximates better the width of the
semi-major axis distribution rather than the FWHM of the surface
density. In fact, I find that $\gamma$ and the disc width are
anticorrelated in the posterior distribution. Hence the values of high
$\rmse$ and low fractional width correspond to high values of
$\gamma$, for which $r_{\max}-r_{\min}$ underestimates the disc FWHM
of the surface density distribution. From the posterior distribution
of the surface density I find that the fractional width of HR~8799
based on its FWHM is constrained between 0.7 and 1.1, i.e. larger than
the value used in this figure and reported in
Table~\ref{tab:results}. Therefore the results for HR~8799's are still
consistent with the limit shown by the dashed line.

As expected, for HD~107146, HD~92945 and HD~206893 the results confirm
the presence of gaps centred at roughly $73$~au. Despite the strong
similarity in the gap radii, their widths and depths seem to be
significantly different. HD~107146's gap is possibly the shallowest
($\sim$50\% empty compared with the peak surface density) and the
widest. In fact, the gap FWHM converges to large values ($\sim90$~au)
that are comparable to the radial extent of the disc. HD~206893's gap
seems to be the deepest and likely fully emptied of material, while
HD~92945's gap is possibly the narrowest (28~au). HR~8799's disc does
not show any evidence of a gap; nevertheless, it is important to note
that it does not cover the range of radii where the other three gaps
are located. In fact, HR~8799~b is on an orbit at 70~au
\citep{Wang2018} that coincides with the gap centers. Hence, the
absence of a gap centred at $\sim70$~au in HR~8799's disc could be
simply due to the three inner planets that would have cleared the disc
inner regions from 10--50~au. On the other hand, AU~Mic's disc does
not extend beyond $\sim50$~au, but a possible gap or inflection point
in the surface density profile is found at 19~au. This is consistent
with the findings by \cite{Daley2019} of the need of emission at
around 11~au in addition to the main belt further out.

The smoothness of the discs inner edges is less constrained as
anticipated in \ref{sec:model}. With the exception of HR~8799, the
inner edges are all consistent with being sharp (i.e. $l_{\rm
  in}/r_{\min}\ll1$) and thus potentially truncated. Note that a sharp
inner edge could still be consistent with high eccentricity particles
further out if these are restricted by a minimum pericentre. In a
scenario where the disc is truncated by an inner planet, the sharpness
of the inner edge is closely related to the planet mass and its
eccentricity \citep{Quillen2006, Chiang2009, Mustill2012,
  Rodigas2014inneredge, Nesvold2015, Tabeshian2016, Regaly2018,
  Dong2020}. Hence an inner planet could be inferred by characterising
how sharp a disc inner edge is
\citep[e.g.][]{Matra2020}. Alternatively, the observed location of the
disc inner edge could be simply due to collisional evolution, i.e. it
marks the radius at which the collisional timescale of the largest
planetesimals is equal to the age of the system
\citep[][]{Kennedy2010, Schuppler2016, Marino2017etacorvi,
  Marino201761vir, Geiler2017}. Interior to this radius, the disc is
strongly depleted due to collisions and is expected to have a surface
density that scales with radius roughly as $r^{7/3}$
\citep{Kennedy2010}. Future work is needed that compare both scenarios
and assess which one resembles better the observed disc inner edges.

Finally, the surface density slopes are all consistent with a flat
surface density of mm-sized grains, except for AU~Mic's disc that is
best fit with an increasing surface density from 10 to 40~au. Flat
surface densities for small and mm-sized grains are expected in broad
discs with large planetesimals that have collisional lifetimes greater
than the system ages \citep{Schuppler2016, Marino201761vir,
  Geiler2017}. For standard strength properties of solids,
\cite{Marino201761vir} found that the surface density slope of
mm-sized grains (i.e. $\gamma$) is $0.6\alpha+0.9$, where $\alpha$ is
the primordial surface density slope of planetesimals. The increasing
surface density slope of mm-sized grains in AU~Mic could indicate that
the primordial surface density slope of planetesimals increases with
radius and has a slope of $\sim1.5$. Alternatively, it could be that
largest planetesimals are already in collisional equilibrium and thus
the surface density follows the $r^{7/3}$ profile described above.

\section{Discussion}
\label{sec:dis}
In this section I discuss the potential origins for the inferred
eccentricities, its implications, and some of the caveats and
limitations of the modelling approach presented in this paper.

\subsection{Self-stirring}
\label{sec:selfstirring}

Here I consider the possibility that the inferred levels of stirring
in HD~107146, HD~92945 and AU~Mic are due to dynamical excitations
between large planetesimals or protoplanets within the disc. It is
well established that planetesimals can stir themselves to levels in
which the velocity dispersion is comparable to the escape velocity of
the largest planetesimals in the disc. Once that level is reached,
collisions tend to dominate over gravitational stirring, halting the
stirring \citep[e.g.][]{Safronov1972, Goldreich2004,
  Schlichting2014}. Therefore, the inferred stirring levels can be
used to find a lower limit on the escape velocity of the most massive
bodies embedded in the disc. For a Rayleigh distribution, the typical
relative velocities between planetesimals is $\sqrt{1.25\es+
  \is}v_\mathrm{Kep}$, where $v_\mathrm{Kep}$ is the local Keplerian
velocity for a circular orbit \citep{Lissauer1993}. The escape
velocity for a spherical object of Diameter $D$ and bulk density
$\rho$ is $\sqrt{2\upi G D^2 \rho /3}$. Equating both, an expression
for the minimum size of the largest planetesimal responsible for
stirring is obtained,
\begin{equation}
  D_{\min}=\sqrt{\frac{3(1.25\es+2h^2)}{2 \upi G \rho}}v_{\rm Kep}. \label{eq:dmin}
\end{equation}
Evaluated at 130~au for a Solar mass star, with a planetesimal bulk
density of 2~g~cm$^{-3}$ and $\rmse=\sqrt{8}h$, $D_{\min}$ becomes
\begin{equation}
  D_{\min} = 605\ \mathrm{km} \left(\frac{\rmse}{0.1} \right) \left(\frac{r}{130\ \mathrm{au}} \right)^{-1} \left(\frac{\rho}{2\ \mathrm{g\ cm^{3}}} \right)^{-1} \left(\frac{M_\star}{1\ M_\odot} \right),
\end{equation}
where $M_{\star}$ is the stellar mass. Consequently, the values found
for HD~107146, HD~92945 and AU~Mic would imply planetesimal diameters
as large as 600, 400 and 500~km, respectively (using the 5\%
percentiles of $\rmse$ in Table \ref{tab:results}, $r_{\max}$ values
in Table~\ref{tab:mcmc}, and stellar masses of 1.0, 0.9 and
0.5~$M_\odot$, respectively). These diameters are equivalent to masses
of 2\%, 0.5\% and 0.8\% of Pluto's mass.

\begin{figure}
  \centering \includegraphics[trim=0.0cm 0.0cm 0.0cm 0.0cm, clip=true,
    width=1.0\columnwidth]{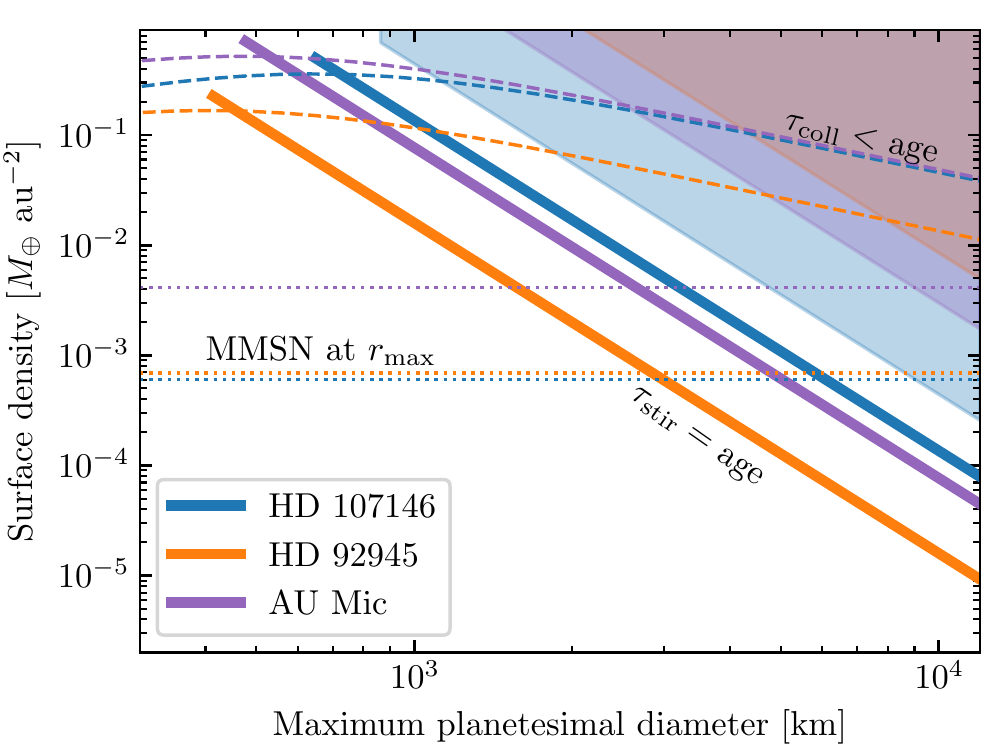}
  \caption{Surface density vs diameter of planetesimals that are
    responsible for stirring the discs around HD~107146 (blue),
    HD~92945 (orange) and AU~Mic (purple). The solid lines represent
    the minimum surface densities to stir the discs within the maximum
    age of these systems. The dashed line represents the maximum
    surface densities that could survive against collisions for the
    maximum estimated age of these systems and minimum inferred level
    of stirring. The blue, orange and purple shaded areas are ruled
    out since the discs would be stirred to high levels that are
    inconsistent with the data. The horizontal dotted lines represent
    the MMSN level at the outer edge of these discs. }
 \label{fig:stir}
\end{figure}

The next simple condition that has to be met is that the level of
stirring could have been reached within the age of the
system. Following the numerical results of \cite{Ida1993} and the more
recent extention of these to a planetesimal size distribution by
\cite{Krivov2018stirring}, the evolution of $\rmse$ as a function of
time $t$ due to viscous stirring in the dispersion-dominated regime is
given by
\begin{equation}
  \rmse=\left( \frac{\pi  \delta C_e \Omega a^2 D_{\max}^3 \rho \Sigma t}{3 M_{\star}^2}\frac{2-\alpha}{3-\alpha} \right)^{1/4}, \label{eq:evst}
\end{equation}
where $\Omega$ is the Keplerian frequency, $\Sigma$ the disc surface
density, $D_{\max}$ the maximum planetesimal size, and $C_e\approx40$
a numerical factor. The factor $\delta$ takes a value between 1 and 2
depending on the eccentricity levels of the largest planetesimals. A
value of unity corresponds to stirrers with zero eccentricities and
inclinations, and a value of 2 to a uniform level of stirring for all
solids in the disc. Equation~\ref{eq:evst} is valid for a planetesimal
mass distribution $dN/dM\propto M^{-\alpha}$. By inverting Equation
~\ref{eq:evst}, an expression for the minimum disc surface density to
reach the level of stirring within the age of the system is obtained
\begin{equation}
    \begin{split}
      \Sigma_{\min}= 0.2\ M_\oplus\ \mathrm{au}^{-2} \left(\frac{\rmse}{0.1}\right)^4 \left(\frac{M_\star}{M_\odot}\right)^{3/2} \left(\frac{t}{100\ \mathrm{Myr}}\right)^{-1}\\ \left(\frac{a}{130\ \mathrm{au}}\right)^{-1/2}\left(\frac{D_{\max}}{1000\ \mathrm{km}}\right)^{-3} \left(\frac{\delta}{2}\right)^{-1},
    \end{split}
\end{equation}
assuming $\alpha=1.6$ as in \cite{Krivov2018stirring}.

Figure~\ref{fig:stir} shows the minimum surface density (solid lines)
for HD~107146, HD~92945 and AU~Mic as a function of maximum
planetesimal diameter to reach $\rmse$ levels of 0.11, 0.07 and 0.06
at the disc outer edges in a timescale of 200, 250 and 21 Myr,
respectively. These timescales correspond roughly to their age upper
limits \citep[][]{Miret-Roig2020, Mesa2021}. Only values above
$D_{\min}$ are shown. These curves show that HD~107146 and AU~Mic
require the largest surface densities to explain the inferred
eccentricity levels at the disc outer edges. Similar to
\cite{Matra2019betapic}, these can be compared with the maximum
surface density that could have survived collisions (assuming the same
level of stirring for the largest planetesimals) that is represented
with dashed lines. This condition does not add further constraints
since the minimum surface density is below the maximum surface density
curve for maximum planetesimal diameters above $D_{\min}$. The
horizontal dotted lines represent the \textit{Minimum Mass Solar
  Nebula} \citep[MMSN, that is equivalent to ($a$/1
  au)$^{-3/2}\ M_\oplus\ \mathrm{au}^{-2}$,][]{Weidenschilling1977mmsn,
  Hayashi1981} at the given disc outer radii of these three
systems. HD~107146 and AU~Mic require similar surface density levels,
although at different radii, while HD92945 requires a surface density
that is an order of magnitude lower. Nevertheless, for a maximum
planetesimal diameter of $\sim1000$~km, the three surface densities
are roughly consistent with the derived dust masses when assuming a
size distribution with a power law index of -3.5. Therefore I conclude
that it is plausible that these discs are self-stirred if the stirrers
have a surface density close to a MMSN and masses similar to Pluto or
larger. Less massive stirrers would require much larger surface
densities, which translate to disc masses in the range 100--1000~\Me.

How and when those massive stirrers formed is an open
question. Simulations by \cite{Kenyon2008} show that forming such
massive bodies through a gas-free planetesimal disc at large radii is
possible, but it can take more than 1 Gyr at these distances, thus
they would have had to form during the gas rich protoplanetary disc
phase. Note that for the high surface densities considered here, the
condition of overlapping feeding zones of the largest planetesimals is
readily met, i.e. there would be enough stirrers across the disc.

Finally, it is still possible that these discs are self-stirred, but
their semi-major axis distributions are even smoother than
considered. In this case the upper limits I derived (8th column in
Table \ref{tab:results}) can be used to set an upper limit on the
planetesimal surface density as a function of the maximum planetesimal
diameter. The shaded regions in \ref{fig:stir} represent the surface
densities and diameters that are ruled out. In these regions the discs
would be stirred to levels that are inconsistent with the smoothness
of their outer edges in timescales shorter than the age lower limits
of these systems (120, 120 and 16 Myr for HD~107146, HD~92945 and
AU~Mic, respectively). In other words, the discs would appear smoother
than observed.



\subsection{Stirring by planets}
\label{sec:planetstirring}
As shown in \S\ref{sec:selfstirring}, stirring by planetesimals is
plausible; however, it requires very high surface densities or
Pluto-sized bodies embedded in the disc. An alternative scenario is
that the smoothness of the outer edges is due to interactions with
planets. These could either stir the whole disc or create additional
hot dynamical populations of particles in analogy to the scattered and
hot populations in the Kuiper belt \citep[see][for a recent
  review]{Morbidelli2020}. I identify multiple possibilities that
could raise the eccentricities and these are discussed below.

\subsubsection{Scattering}
First, a planet formed close to the initial disc inner edge could have
undergone planetesimal-driven outward migration while scattering
planetesimals onto high eccentricity orbits with larger semi-major
axes, similar to the proposed Neptune migration into the Kuiper belt
\citep[e.g.][]{Malhotra1993, Malhotra1995, Nesvorny2015}. This
scattered population of planetesimals would be characterised by
pericentres close to the disc inner edge, or further out if they
become detached (as in the detached or fossilised scattered population
in the Kuiper belt). This population would also have apocentres
greater than the inferred disc outer edges such that their
distribution smooths the surface density in the outer regions. This
scenario could be consistent with HR~8799's properties, especially
since a scattered population could be caused by the four known giant
planets as they cleared the regions near their current orbits and
possibly an even wider region if they migrated to their current orbits
\citep{Read2018, Gozdziewski2018, Gozdziewski2020}. This idea is
consistent with the two population model proposed by \cite{Geiler2019}
to explain Herschel FIR and ALMA mm wavelength data.

Alternatively, a massive planet in the middle of a disc could clear a
gap as seen in HD~107146, HD~92945, HD~206893 and potentially AU~Mic
by scattering material forming a scattered component. However, N-body
simulations by \citet[see their Fig. 5]{Marino2018hd107} did not
produce a noticeably smooth outer edge with planet masses below
90~\Me. Those simulations nevertheless ignored disc self-gravity,
planet migration or more massive planets, and thus this scenario
cannot be ruled-out yet. Note that most of the literature on scattered
components has focused on the Solar System, and more general studies
are needed to study how a scattered component could smooth the disc
outer edge and the effect of collisions \citep[as
  in][]{Wyatt2010}. Lastly, it is also possible that a low mass planet
formed near the disc outer edge could undergo planetesimal-driven
inward migration, crossing and stirring the entire disc within the age
of these systems \citep{Kirsh2009}.

\subsubsection{Mean motion resonances}

Second, a planet embedded in the disc or at its inner edge could have
trapped planetesimals in resonances and increase their eccentricities
while it was migrating \citep[e.g.][]{Levison2003kb,
  Wyatt2003}. However, for a wide disc a planet near the inner edge
would hardly excite the eccentricity of particles near the outer edge
via mean-motion resonances, and scattering is more likely as also
concluded by \cite{Matra2019betapic} for the case of $\beta$~Pic.

\subsubsection{Secular interactions}

Third, in analogy to the Nice model \citep{Tsiganis2005, Gomes2005,
  Nesvorny2012}, a dynamical instability between inner planets in
these systems could have raise their eccentricities and thus excited
the orbits of planetesimals via secular interactions
\citep{Mustill2009}. This scenario would smooth the disc outer edges
and could also produce half emptied gaps
\citep[][]{Pearce2015doublering}. However, this scenario is less
likely to explain HD~107146's axisymmetric gap and HD~206893's fully
emptied gap.

\subsubsection{Secular resonances}
Finally, secular resonances could also excite the eccentricity of
solids in the disc at specific locations if one or multiple planets
reside interior to the disc \citep{Zheng2017, Yelverton2018,
  Sefilian2020}. The location of these resonances can shift in time as
the disc losses mass due to collisional evolution or as planets
migrate, thus exciting the eccentricity of planetesimals over a wide
range of semi-major axis.

Therefore, I conclude that a single or multiple planets could be
responsible for the inferred excitation levels via scattering, an
instability and subsequent secular interactions with the disc, or
sweeping secular resonances.

\subsection{A primordial smooth outer edge in the semi-major axis distribution}

Throughout this paper I have assumed that the observed smoothness of
the disc outer edges is due to the orbital eccentricities and that the
distribution of semi-major axes has a sharp drop or at least sharper
than the observed levels. If this is truly the case depends strongly
on how planetesimals are formed and how this varies as a function of
radius. If planetesimals at tens of au are formed via streaming
instability \citep{Youdin2005, Johansen2007} triggered in radial dust
traps in protoplanetary disks \citep[e.g.][]{Pinilla2012a,
  Pinilla2020a, Dullemond2018dsharp, Stammler2019}, the resulting
planetesimal disc could have a smooth or sharp outer edge depending on
how radially concentrated were dust particles and how those traps
evolve in time \citep[e.g.][Miller et al. in prep]{Lenz2019,
  Shibaike2020}. Alternatively, streaming instability could be
triggered while the gas surface density drops due to photoevaporation
\citep{Throop2005, Carrera2017, Ercolano2017}, in which case the
planetesimal distribution could have a steep outer edge if the dust
distribution had also a steep outer edge. However, more recent models
of dust evolution in photoevaporating discs show that the dust-to-gas
ratio is weakly affected due to efficient radial drift
\citep[e.g.][]{Sellek2020}, and thus photoevaporation might not
trigger the streaming instability. Therefore, dust traps might be
necessary to overcome radial drift even when considering
photoevaporation.

In \S\ref{sec:retrieve} I argue that, despite the fact that the
smoothness of the surface density could be a consequence of either the
semi-major axis or eccentricity distributions (or both), it can
nevertheless set an upper limit on the eccentricity
distribution. Similarly, by assuming eccentricities are zero the
derived smoothness values constrain the maximum smoothness in the
semi-major axis distribution. In this way, the smoothness could be
used to test and constrain planetesimal formation models in the
future.





\subsection{$\rmse$ varying with a}

Another strong assumption through this paper is that the distributions
of eccentricities and semi-major axes are independent and that the
former follows a Rayleigh distribution. There are multiple scenarios
in which this is not the case. For example, in a disc of particles
being scattered by an inner planet the eccentricities are strongly
correlated with the semi-major axis of particles, with eccentricities
approximating $1-q/a$, where $q$ is the common pericenter near the
planet's orbit. In this case the disc outer edge should follow
approximately a power law distribution with a power law index of -3
\citep{Duncan1987, Marino2018scat}, although collisional evolution
tends to flatten the surface density distribution over time
\citep{Wyatt2010}. In order to test what would be the result of
fitting the parametric model described by Equation \ref{eq:tanh} to a
scattered disc, I fit this model to several simulated discs with power
law outer edges declining as $r^{-3}$. I find that
$l_\mathrm{out}/r_{\max}$ converges to values $>0.5$ for wide discs
and values closer to 1 for narrower discs, thus consistent with the
results for HD~206893 and HR~8799. Note that this is not proof of
those discs having a scattered component; nevertheless, it shows that
the results of very smooth edges are consistent with that
possibility. Future work comparing scattered disc and alternative
models to the data could provide more support to this hypothesis.

Another possibility is that the eccentricities do follow a Rayleigh
distribution, but the level of dynamical excitation or $\rmse$ varies
as a function of semi-major axis. This might be the most natural
outcome of self-stirring or stirring via secular interactions with a
planet. In a self-stirring scenario, $\rmse$ should vary as $(a^{1/2}
\Sigma D_{\max})^{1/4}$ (see Equation~\ref{eq:evst}). Thus if either
$\Sigma$ or $D_{\max}$ decreased with semi-major axis more steeply
than $a^{-1/2}$ \citep[e.g.][]{Kenyon2008, Lenz2019, Klahr2020},
$\rmse$ should decrease with radius. Conversely, if stirring has
reached its equilibrium (i.e. the relative velocities are roughly the
same as the escape velocity of the largest planetesimals), the
eccentricities could increase with radius for a constant maximum
planetesimal size (see Equation \ref{eq:dmin}). If eccentricities are
instead excited via secular interactions with an inner planet, $\rmse$
should decrease with semi-major axis as $1/a$
\citep[e.g.][]{Mustill2009}. In either of these cases, the estimates
of $\rmse$ are still valid near the outer edge of the disc.

\subsection{Halos observed at mm-wavelengths}

Finally, evidence of high levels of stirring could also exist in other
systems such as HD~181327, HD~61005 and HD~32297. Those systems were
found to posses halos or extended components of mm-grains beyond their
main belt \citep{Marino2016, MacGregor2018}. Such halos could be
indicative of scattered discs. While HD~61005 and HD~32297 are seen
edge-on and thus their radial profiles are hard to disentangle
directly from the images, HD~181327 is seen close to face-on. Its main
belt is narrow (width of 21~au) and centered at 80~au \citep[after
  correcting by the Gaia DR3 distance of 48 pc,][]{Gaiadr3}, but it
displays low level emission out to $\sim200$~au \citep{Marino2016}. If
this halo was produced by particles with pericentres at the belt inner
edge (70~au) and apocentres as large as 200~au, that would imply
eccentricities as high as 0.5. On the other hand, HD~61005 and
HD~32297 seemed to have smooth outer edges instead of an instant drop;
however, \cite{MacGregor2018} found their outer edges had power law
slope values of $\sim-6$, significantly steeper than expected for a
simple scattered disc. Further observations and modelling are
necessary to assess if these halos are consistent with scattered discs
or not.


\section{Conclusions}
\label{sec:con}

In this paper, I have shown how the stirring levels of a planetesimal
disc can influence its appearance (\S\ref{sec:model}). Namely, the
higher the dispersion of eccentricities ($e$) is, the smoother the
disc surface density will be. This effect is most noticeable at the
disc outer edge since features are smoothed by a length scale $\sim
r\rmse$ if eccentricities levels are uniform across the disc. I found
that a sharp edge in the distribution of semi-major axes, translates
to a smooth edge in the surface density radial profile that bears
strong similarities with a hyperbolic tangent function (Equation
\ref{eq:tanh}).


By parametrising the disc surface density as a power law distribution
with edges following hyperbolic tangents, the eccentricity levels in a
disc can be retrieved. In \S\ref{sec:retrieve} I found that when
$\rmse$ is smaller than the disc fractional width, $\rmse$ is well
approximated by $\approx 1.2 l_{\rm out}/r_{\max}$, where $l_{\rm
  out}$ and $r_{\max}$ are parameters that determine the smoothness
length and location of the disc outer edge, respectively. These
relations assume, however, that the semi-major axis distribution has
sharp edges, which might be unrealistic. It nevertheless represents a
limit case in which the smoothness is only due to stirring. A smoother
semi-major axis distribution would result in even smoother edges,
therefore $1.2 l_{\rm out}/r_{\max}$ provides an upper limit on
$\rmse$ near the disc outer edge.

There are particular scenarios in which the semi-major axis
distribution could become smoother. Such is the case of a self-stirred
disc, where the increase in eccentricities due to close encounters
with massive planetesimals should also lead to a diffusion in
semi-major axis. I explored this case in \S\ref{sec:smootha}, finding
that $\rmse\approx 0.77 l_{\rm out}/r_{\max}$, i.e. the necessary
eccentricity levels to explain the outer edge smoothness are lower.

In \S\ref{sec:application} these findings are applied to real
observations by fitting the parametric model to the ALMA data of five
discs that are well resolved: HD~107146, HD~92945, HD~206893, HR~8799
and AU~Mic. The five discs require $l_{\rm out}/r_{\max}>0$, i.e. the
shape of their outer edges is at least marginally resolved. I find
that HD~107146, HD~92945 and AU~Mic require lower eccentricities
(i.e. have sharper outer edges) compared with HD~206893 and
HR~8799. Assuming a self-stirring scenario, the former group requires
$\rmse$ values of $0.121\pm0.05$, $0.15^{+0.07}_{-0.05}$ and
$0.10\pm0.02$, respectively. Interestingly, the value of $\rmse$ and
scale height derived for HD~92945 are consistent with a scenario in
which the radial and vertical stirring are in equilibrium. On the
other hand, HD~206893 and HR~8799 require much smoother outer edges
with $\rmse$ values $\sim0.5$ if their semi-major axis distribution
had sharp edges. Such high eccentricity levels would be more
consistent with stronger interactions with massive planets. It is
perhaps not a coincidence that these systems also host massive inner
companions. Such companions could have migrated through planetesimal
driven migration in the past, scattering a large population of
planetesimals raising their eccentricities and semi-major axes.


Using the derived stirring levels, in \S\ref{sec:selfstirring} I
explored the possibility that the discs around HD~107146, HD~92945 and
AU~Mic are self-stirred. I found that to be able to stir the disc to
the inferred levels, planetesimals of at least 400~km in diameter
should be present. Given the age of these systems, I also constrained
how high should the planetesimal surface density be to reach the
observed levels within the age of these systems. I found that if
planetesimals are smaller than Pluto the required surface densities
are significantly higher than a MMSN and vice versa. While this
finding favours the presence of Pluto-size objects in the disc, the
formation of such massive bodies through collisions of smaller
planetesimals is too slow to explain their presence. This problem
could be overcome if these massive bodies are formed already in
protoplanetary discs.


Alternatively, the discs could have been stirred by planets
(\S\ref{sec:planetstirring}). I identified four mechanisms that could
stir the disc smoothing the outer edges, and conclude that scattering,
secular interactions and secular resonances are the most promising to
stir these discs via planet-disc interactions. While only two of these
five systems have known companions (HD~206893 and HR~8799), the known
gaps in the rest suggests the presence of unseen lower mass companions
interacting with the discs. The presence of more massive companions
around HD~206893 and HR~8799 could lead to a higher excitation of
their discs, explaining their very smooth outer edges. In fact, their
smoothness levels derived from the fits are consistent with typical
scattered disc profiles.

Finally, I hope the work presented here serves as a future guide on
how characterize how sharp or smooth are the edges of planetesimal
discs. While these edges are set by both the semi-major axis and
eccentricity distribution and thus it is a degenerate problem, the
smoothness of the surface density provides an upper limit for both the
smoothness of the semi-major axis distribution and dispersion of
eccentricities. Coupling this analysis with vertically resolved
observations could significantly help to lift degeneracies and test
the different assumptions made in this work.

\section*{Acknowledgements}

I would like to thank the anonymous referee for a very constructive
review that improved the quality and clarity of this paper. I would
also like to thank Luca Matr\`a for providing AU Mic's calibrated data
set, Virginie Faramaz and Mark Booth for providing HR~8799 calibrated
data sets, and Alexander Krivov for noticing an error in equation 9
that has been corrected before publication. Sebastian Marino is
supported by a Junior Research Fellowship from Jesus College,
Cambridge.  This paper makes use of the following ALMA data:
ADS/JAO.ALMA\#2016.1.00104.S, ADS/JAO.ALMA\#2016.1.00195.S,
ADS/JAO.ALMA\#2019.1.00189.S, ADS/JAO.ALMA\#2017.1.00828.S,
ADS/JAO.ALMA\#2017.1.00825.S, ADS/JAO.ALMA\#2012.1.00198.S,
ADS/JAO.ALMA\#2012.1.00482.S, ADS/JAO.ALMA\#2016.1.00907.S. ALMA is a
partnership of ESO (representing its member states), NSF (USA) and
NINS (Japan), together with NRC (Canada), MOST and ASIAA (Taiwan), and
KASI (Republic of Korea), in cooperation with the Republic of
Chile. The Joint ALMA Observatory is operated by ESO, AUI/NRAO and
NAOJ.

\section*{Data availability}
The data underlying this article will be shared on reasonable request
to the corresponding author. The ALMA data is publicly available and
can be queried and downloaded directly from the ALMA archive at
https://almascience.nrao.edu/asax/.




\bibliographystyle{mnras}
\bibliography{SM_pformation} 



\appendix

\section{Data and visibility fitting details}
\label{appendix:fit}

\begin{table*}
  \centering
  \caption{Best fit values of the disc parameters of HD~107146,
    HD~92945, HD~206893, AU~Mic and HR~8799. The best fit values
    correspond to the median, with uncertainties based on the 16th and
    84th percentiles of the marginalised distributions. The 2$^{\rm
      nd}$ column corresponds to the dust mass. The 3$^{\rm
      rd}$--7$^{\rm th}$ columns correspond to the parameters that
    define the surface density according to
    Equation~\ref{eq:tanh}. The 8$^{\rm th}$, 9$^{\rm th}$ and
    10$^{\rm th}$ columns correspond to the gap central radius, its
    FWHM and its depth (with a value of 1 meaning a completely empty
    gap at its center), respectively. The 11$^{\rm th}$ and 12$^{\rm
      th}$ column show the disc inclination from a face-on orientation
    and its position angle. The 13$^{\rm th}$ column displays the disc
    vertical aspect ratio (constant across the disc), assuming a
    Gaussian vertical mass distribution with a standard deviation of
    $hr$. The final column shows the spectral index of the disc
    emission. This parameter was fixed to 0.05 for HD~107146,
    HD~206893 and HR~8799, and left as a free parameter for HD~92945
    and AU~Mic. }
  \label{tab:mcmc}
  \begin{adjustbox}{max width=1.0\textwidth}
    \begin{tabular}{lccccccccccccc} 
  \hline
  \hline
  System & $M_{\rm dust}$ & $r_{\min}$ & $r_{\max}$ & $\gamma$ & $l_{\rm in}$ & $l_{\rm out} $ & $\rgap$ & $\wgap$ & $\dgap$ & inc & PA & h & $\alpha_{\rm mm}$\\
  & [$M_{\oplus}$] & [au] &   [au]       &          & [au]        & [au]         & [au]     & [au] &    & [deg] & [deg] &  \\
    (1) & (2) & (3) & (4) & (5) & (6) & (7) & (8) & (9) & (10) & (11) & (12) & (13) & (14)  \\

  \hline
   HD~107146 & $0.224_{-0.003}^{+0.003} $ & $38.3_{-0.6}^{+0.6} $ & $137.8_{-2.0}^{+1.8} $ & $0.0_{-0.2}^{+0.2} $ & $4.4_{-1.2}^{+1.1} $ & $21.8_{-0.8}^{+0.8} $ & $72.2_{-1.5}^{+1.7} $ & $92.9_{-14.7}^{+14.7} $ & $0.77_{-0.06}^{+0.05} $ & $19.9_{-0.6}^{+0.6} $ & $153.3_{-1.5}^{+1.6} $ &  & $2.55_{-0.04}^{+0.04}$\\
  HD~92945 & $0.048_{-0.002}^{+0.002} $ & $50.3_{-2.4}^{+3.0} $ & $128.3_{-14.4}^{+7.3} $ & $-0.6_{-0.7}^{+0.6} $ & $6.1_{-2.8}^{+2.3} $ & $24.9_{-7.6}^{+8.0} $ & $73.7_{-1.7}^{+1.8} $ & $27.6_{-9.9}^{+8.7} $ & $0.73_{-0.10}^{+0.08} $ & $65.4_{-0.6}^{+0.6} $ & $100.0_{-0.6}^{+0.6} $ & $0.043_{-0.014}^{+0.014} $ \\
  HD~206893  & $0.029_{-0.002}^{+0.002} $ & $26.2_{-16.2}^{+13.1} $ & $116.9_{-16.3}^{+21.3} $ & $1.0_{-0.7}^{+0.5} $ & $17.9_{-12.5}^{+19.8} $ & $46.7_{-6.0}^{+5.6} $ & $73.0_{-3.5}^{+3.2} $ & $31.8_{-5.7}^{+7.5} $ & $0.90_{-0.10}^{+0.07} $ & $40.0_{-2.6}^{+2.4} $ & $61.7_{-3.8}^{+3.7} $ & & $2.45^{+0.15}_{-0.17} $\\
  HR~8799 & $0.132_{-0.013}^{+0.013} $ & $126.5_{-23.4}^{+21.4} $ & $271.0_{-93.6}^{+75.7} $ & $1.0_{-1.1}^{+1.6} $ & $40.8_{-15.5}^{+12.6} $ & $113.6_{-19.3}^{+21.3} $ & & & &   $31.2_{-3.1}^{+2.5} $ & $52.3_{-5.5}^{+5.2}$ & &  $2.49^{+0.24}_{-0.22}$ \\
   AU~Mic & $0.0143_{-0.0002}^{+0.0003} $ & $13.2_{-6.4}^{+4.4} $ & $38.5_{-1.8}^{+1.4} $ & $1.8_{-0.9}^{+1.2} $ & $8.1_{-5.7}^{+5.1} $ & $5.2_{-1.1}^{+0.9} $ & $18.8_{-1.2}^{+1.4} $ & $10.2_{-4.1}^{+4.6} $ & $0.92_{-0.21}^{+0.06} $ & $88.2_{-0.1}^{+0.1} $ & $128.5_{-0.1}^{+0.1} $ & $0.015_{-0.003}^{+0.003} $ \\
  \hline
  
  \end{tabular}
  
  \end{adjustbox}
\end{table*}

The best fit values for the disc parameters of the five systems are
presented in Table \ref{tab:mcmc}. Below I describe the data used in
the analysis and a few particularities in the models used to fit each
system.

\subsubsection{HD~107146}
The ALMA data of HD~107146 used in this paper is described in
\cite{Marino2018hd107}, and corresponds to band 7 observations as part
of the project 2016.1.00104.S (PI: S. Marino), and band 6 observations
as part of the project 2016.1.00195.S (PI: J. Carpenter). In addition,
I use new data in band 7 from the project 2019.1.00189.S (PI:
S. Marino). The new data were taken in December 2019 and corresponds
to a single compact antenna configuration (resolution of
$\sim1\arcsec$), thus providing marginal improvement over previous
observations. The data will be described in more detail in a future
paper after completion of the observations, which will include higher
resolution observations to be taken once ALMA observations resume
after the 2020 shutdown due to the COVID-19 pandemic.

The new band 7 data revealed that the clump identified in
\cite{Marino2018hd107} that could have corresponded to its possible
warm asteroid belt \citep{Morales2011, Kennedy2014}, is not co-moving
with HD~107146, and thus it is most likely a background galaxy. In
order to account for this background source in the modelling, I add to
the model images an additional non co-moving component. This
additional component has a surface brightness distribution
parametrised as a 2D elliptical Gaussian and with a relative position
to HD~107146 that changes according to HD~107146's proper motion and
epoch of observation. This background source is fit in parallel with
the disc emission. I find that the background source is best fit by an
offset from HD~107146 in December 2019 of $-0\farcs01\pm0\farcs03$ in
RA and $0\farcs09\pm0\farcs02$ in Dec. Its flux at 0.88 mm is
$0.79\pm0.07$ mJy and has a spectral index of $3.2\pm0.2$, hence
consistent with a background sub-mm galaxy. Its standard deviation
along its major axis is $0\farcs64\pm0\farcs05$ and along its minor
axis $0\farcs53\pm0\farcs04$ (PA of $94\degr\pm12\degr$), thus it is
roughly circular.

\subsubsection{HD~92945}
The ALMA data of HD~92945 used in this paper is described in
\cite{Marino2019}, and corresponds to band 7 observations as part of
the project 2016.1.00104.S (PI: S. Marino). Because the star is
marginally detected, I leave its flux as a free parameter and find a
best fit value of $35\pm15$~$\mu$Jy.

\subsubsection{HD~206893}
The ALMA data of HD~206893 used in this paper corresponds to band 7
data described in \cite{Marino2020hd206} as part of the project
2017.1.00828.S (PI: A. Zurlo), and band 6 data described in
\citep{Nederlander2021} as part of the project 2017.1.00825.S (PI:
A. M. Hughes). Since this system has a massive companion on an orbit
near 10 au that would have cleared any debris near its orbit, the
surface density is forced to zero within 14~au. This is roughly the
minimum outer extent of the companion chaotic zone if on a circular
orbit.

\subsubsection{HR~8799}
The ALMA data of HR~8799 used in this paper corresponds to band 6 data
described in \cite{Booth2016} as part of the project 2012.1.00482.S
(PI:A. Jordan), and band 7 data described in Faramaz et
al. (submitted) as part of the project 2016.1.00907.S (PI:
V. Faramaz). Channels with CO emission in both data sets were
flagged. Since this system has massive companions interior to the disc
that would have cleared any debris, the surface density is forced to
zero within 70~au. This is roughly the semi-major axis of the outer
most planet. Because HR~8799 also presents a bright background galaxy
that overlays with the disc (Faramaz et al. submitted), I include an
additional component in the modelling using the same approach as for
HD~107146. The background source is best fit by an offset from HR~8799
in June 2018 of $-1\farcs28\pm0\farcs05$ in RA and
$2\farcs34\pm0\farcs05$ in Dec. Its flux at 0.88 mm is $0.31\pm0.02$
mJy and $0.06\pm0.02$ mJy at 1.3 mm. The central flux (stellar
photosphere plus any other additional emission) is also left as a free
parameter, finding a best fit of $71\pm10$ $\mu$Jy at 0.88 mm and
$39\pm16$ $\mu$Jy at 1.3mm.

\begin{figure}
  \centering \includegraphics[trim=0.0cm 0.0cm 0.0cm 0.0cm, clip=true,
    width=1.0\columnwidth]{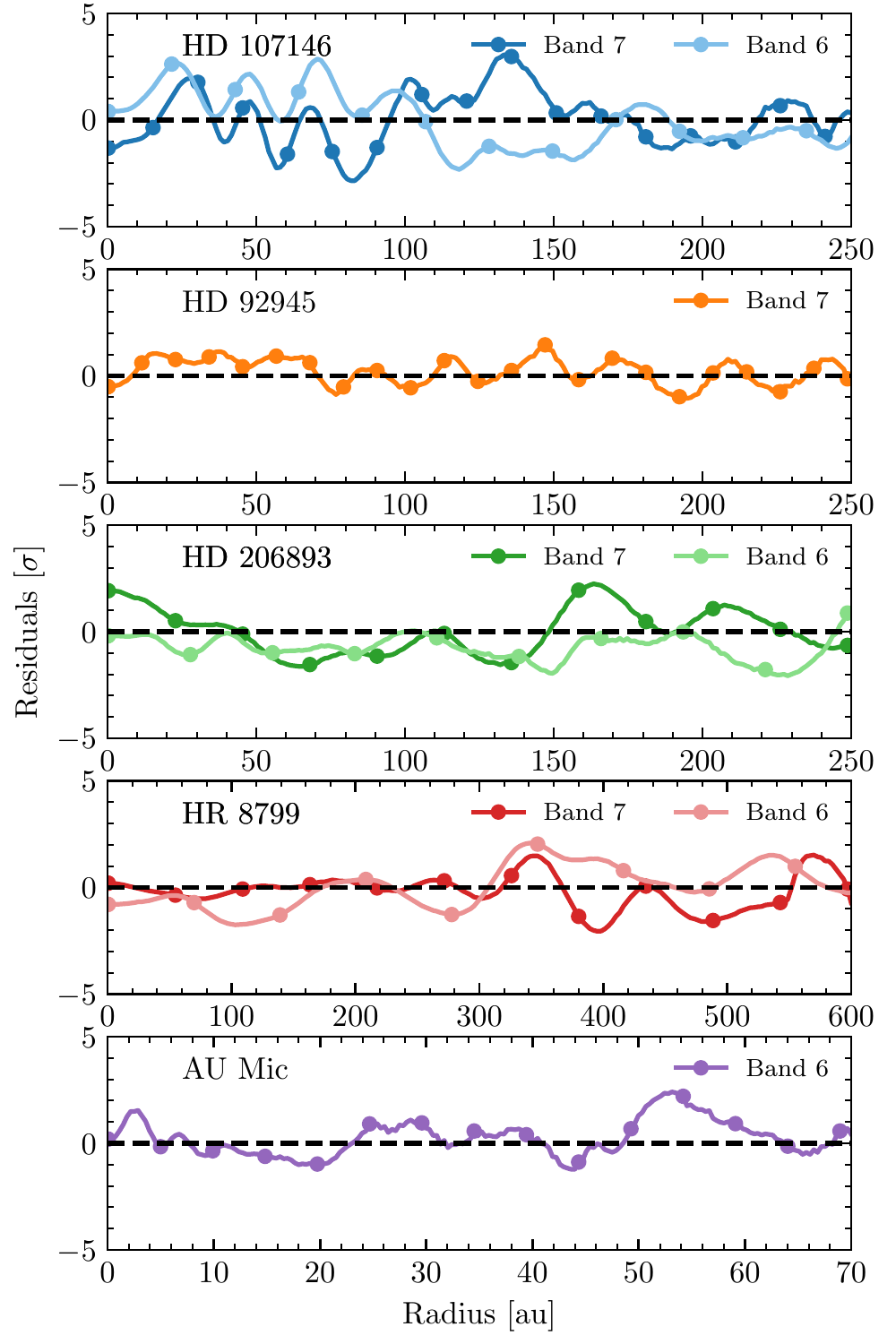}
  \caption{Radial profile of the azimuthally averaged residuals after
    subtracting the best fit models. The y axis represents the
    residuals in units of the local uncertainty after azimuthal
    averaging, while the x axis represents the deprojected
    radius. Since AU~Mic's disc is edge-on, the profile is obtained
    simply as a cut through the disc major axis. The spacing between
    the dots represent the beam major axis for each image.}
 \label{fig:residuals}
\end{figure}

\subsubsection{AU Mic}
The ALMA data of AU~Mic used in this paper corresponds to band 6 data
described in \cite{Daley2019} as part of the project 2012.1.00198.S
(PI: A. M. Hughes). I use a reduced data provided by L. Matr\`a which
will be presented as part of the REASONS survey \citep[][Matr\`a et
  al. in prep]{Sepulveda2019}. This data set was flagged at epochs
when strong flares were present, and its three individual observations
re-centred to the stellar position via visibility modelling. Because
the star is strongly detected, I leave its flux as a free parameter
and find a best fit value of $0.26\pm0.02$ mJy.

Figure \ref{fig:residuals} shows the radial profiles of the
deprojected and azimuthally averaged residual maps in units of the
local uncertainty. These maps correspond dirty maps of the residuals,
obtained with CASA using Briggs weighting and a robust parameter of
2.0. For HD~206893 and HR~8799 band 7 data, an additional uvtaper of
$0\farcs4$ and $0\farcs8$ was used, respectively, to increase the
sensitivity per beam. There are no residuals above or below $3\sigma$
in any of the profiles, meaning the best fits are able to reproduce
all the relevant radial features in the data. Since AU~Mic's disc is
edge on, deprojecting its disc emission is not possible without
modelling. Thus the presented residual radial profile for this source
is simply a cut through the disc major axis.




\bsp	
\label{lastpage}
\end{document}